\documentclass[longauth]{aa}

\usepackage{graphicx}
\usepackage{txfonts}
\usepackage{natbib}
\usepackage[colorlinks,urlcolor=cyan,citecolor=blue,linkcolor=blue]{hyperref}
\usepackage{amssymb,amsmath}
\usepackage{threeparttable}
\usepackage{array}
\usepackage{booktabs}
\usepackage{multirow}
\newcommand{\tls}{{\fontfamily{pcr}\selectfont tls}\,}
\newcommand{\ellc}{{\fontfamily{pcr}\selectfont ellc}\,}
\newcommand{\emcee}{{\fontfamily{pcr}\selectfont emcee}\,}
\newcommand{\dynesty}{{\fontfamily{pcr}\selectfont dynesty}\,}
\newcommand{\allesfitter}{{\fontfamily{pcr}\selectfont allesfitter}\,}
\newcommand{\sherlock}{{\fontfamily{pcr}\selectfont SHERLOCK}\,}
\newcommand{\matrixtk}{{\fontfamily{pcr}\selectfont MATRIX}\,}
\newcommand{\wotan}{{\fontfamily{pcr}\selectfont wotan}\,}
\newcommand{\forecaster}{{\fontfamily{pcr}\selectfont forecaster}\,}
\newcommand{\lightkurve}{{\fontfamily{pcr}\selectfont lightkurve}\,}
\newcommand{\celerite}{{\fontfamily{pcr}\selectfont celerite}\,}
\newcommand{\triceratops}{{\fontfamily{pcr}\selectfont TRICERATOPS}\,}

\usepackage{tikz,xcolor,hyperref}

\definecolor{lime}{HTML}{A6CE39}
\DeclareRobustCommand{\orcidicon}{%
	\hspace{-1.5mm}
	\begin{tikzpicture}
	\draw[lime, fill=lime] (0,0) 
	circle [radius=0.16] 
	node[white] {{\fontfamily{qag}\selectfont \tiny ID}};
	\draw[white, fill=white] (-0.0625,0.095) 
	circle [radius=0.007];
	\end{tikzpicture}
	\hspace{-2.5mm}
}

\foreach \x in {A, ..., Z}{%
	\expandafter\xdef\csname orcid\x\endcsname{\noexpand\href{https://orcid.org/\csname orcidauthor\x\endcsname}{\noexpand\orcidicon}}
}

\foreach \x in {a, ..., z}{%
	\expandafter\xdef\csname orcid\x\endcsname{\noexpand\href{https://orcid.org/\csname orcidauthor\x\endcsname}{\noexpand\orcidicon}}
}

\foreach \x in {a, ..., z}{%
	\expandafter\xdef\csname orcid\x\x\endcsname{\noexpand\href{https://orcid.org/\csname orcidauthor\x\x\endcsname}{\noexpand\orcidicon}}
}

\begin{document}

\title{A super-Earth and a mini-Neptune near the 2:1 MMR straddling the radius valley around the nearby mid-M dwarf TOI-2096}
\titlerunning{The TOI-2096 planetary system}

\author{
F.J.~Pozuelos\orcidA{}\inst{\ref{aru_liege},\ref{iaa},\ref{star_liege}}
\and M.~Timmermans\inst{\ref{aru_liege}}
\and B.V.~Rackham\orcidJ{}\inst{\ref{miteaps},\ref{mitkavli},\thanks{51 Pegasi b Fellow}} 
\and L.J.~Garcia\orcidW{}\inst{\ref{aru_liege}}
\and A.J.~Burgasser\orcidI{}\inst{\ref{san_diego}} 
\and S.R.~Kane\orcidj{}\inst{\ref{riverside}} 
\and M.N.~G\"unther\orcidB{}\inst{\ref{estec},\thanks{ESA Research Fellow}} 
\and K.G.~Stassun\orcidK{}\inst{\ref{vanderbilt}} 
\and V.~Van~Grootel\orcidL{}\inst{\ref{star_liege}} 
\and M.~D\'evora-Pajares\orcidt{}\inst{\ref{ugr}} 
\and R.~Luque\orcidu{}\inst{\ref{chicago},\ref{iaa}}
\and B.~Edwards\orcidv{}\inst{\ref{cea},\ref{clondon},\thanks{Paris Region Fellow, Marie Sklodowska-Curie Action}} 
\and P.~Niraula\orcidO{}\inst{\ref{miteaps}}
\and N.~Schanche\inst{\ref{unibe},\ref{maryland},\ref{nasa_goddard}} 
\and R.D.~Wells\orcidw{}\inst{\ref{unibe}} 
\and E.~Ducrot\orcidH{}\inst{\ref{cea},\thanks{Paris Region Fellow, Marie Sklodowska-Curie Action}}
\and S.~Howell\orcidN{}\inst{\ref{nasa_ames}}
\and D.~Sebastian\orcidP{}\inst{\ref{birmingham}} 
\and K.~Barkaoui\orcidQ{}\inst{\ref{aru_liege},\ref{miteaps},\ref{iac}}
\and W.~Waalkes\inst{\ref{boulder}} 
\and C.~Cadieux\orcidp{}\inst{\ref{montreal}} 
\and R.~Doyon\inst{\ref{montreal},\ref{montreal2}}
\and R.~P.~Boyle\inst{\ref{vatican}} 
\and J.~Dietrich\inst{\ref{arizona}} 
\and A.~Burdanov\orcidR{}\inst{\ref{miteaps}}
\and L.~Delrez\orcidb{}\inst{\ref{aru_liege},\ref{star_liege}}
\and B.-O.~Demory\orcidS{}\inst{\ref{unibe}}
\and J.~de~Wit\inst{\ref{miteaps}}
\and G.~Dransfield\orcidT{}\inst{\ref{birmingham}} 
\and M.~Gillon\orcidC{}\inst{\ref{aru_liege}}
\and Y.~G\'omez~Maqueo~Chew\orcidY{}\inst{\ref{ciudad}} 
\and M.J.~Hooton\orcidi{}\inst{\ref{cavendish}}
\and E.~Jehin\orcida{}\inst{\ref{star_liege}} 
\and C.A.~Murray\orcidG{}\inst{\ref{boulder},\ref{cavendish}}
\and P.P.~Pedersen\inst{\ref{cavendish}}
\and D.~Queloz\orcidd{}\inst{\ref{cavendish}} 
\and S.J.~Thompson\orcidh{}\inst{\ref{cavendish}}   
\and A.H.M.J.~Triaud\orcidg{}\inst{\ref{birmingham}}   
\and S.~Z\'u\~niga-Fern\'andez\orcids{}\inst{\ref{aru_liege}} 
\and K.A.~Collins\inst{\ref{cfa}} 
\and M.M~Fausnaugh\orcidU{}\inst{\ref{mit2}}
\and C.~Hedges\inst{\ref{nasa_goddard}} 
\and K.M.~Hesse\inst{\ref{mit2}} 
\and J.~M.~Jenkins\orcidE{}\inst{\ref{nasa_ames}} 
\and M.~Kunimoto\inst{\ref{mit2}} 
\and D.W.~Latham\orcidc{}\inst{\ref{cfa}} 
\and A.~Shporer\inst{\ref{mit2}} 
\and E.B.~Ting\orcidD{}\inst{\ref{nasa_ames}} 
\and G.~Torres\inst{\ref{cfa}} 
\and P.~Amado\orcidk{}\inst{\ref{iaa}} 
\and J.R.~Rod\'on\orcidn{}\inst{\ref{iaa}} 
\and C.~Rodr\'iguez-L\'opez\orcidl{}\inst{\ref{iaa}} 
\and J.C.~Su\'arez\orcidm{}\inst{\ref{ugr}} 
\and R.~Alonso\inst{\ref{iac},\ref{la_laguna}}
\and Z.~Benkhaldoun\inst{\ref{tn}} 
\and Z.~K.~Berta-Thompson\orcido{}\inst{\ref{boulder}} 
\and P.~Chinchilla\inst{\ref{aru_liege},\ref{iac}}
\and M.~Ghachoui\inst{\ref{aru_liege},\ref{tn}}
\and M.A.~G\'omez-Mu\~noz\orcidX{}\inst{\ref{ciudad}} 
\and R. Rebolo\orcidx{}\inst{\ref{iac},\ref{la_laguna}}
\and L.~Sabin\orcide{}\inst{\ref{ciudad}} 
\and U.~Schroffenegger\inst{\ref{unibe}} 
\and E.~Furlan\orcidV{}\inst{\ref{nasa_exo}} 
\and C.~Gnilka\inst{\ref{nasa_ames},\ref{nasa_exo}} 
\and K.~Lester\inst{\ref{nasa_ames}}
\and N.~Scott\inst{\ref{chara}} 
\and C.~Aganze\orcidM{}\inst{\ref{san_diego}}
\and R.~Gerasimov\inst{\ref{san_diego}}
\and C.~Hsu\orcidF{}\inst{\ref{North}}  
\and C.~Theissen\orcidf{}\inst{\ref{san_diego},\thanks{NASA Sagan Fellow}} 
\and D.~Apai\orcidr{}\inst{\ref{arizona},\ref{arizona2}} 
\and W.~P.~Chen\inst{\ref{taiwan}}
\and P.~Gabor\inst{\ref{vatican}}
\and T.~Henning\inst{\ref{max}}
\and L.~Mancini\inst{\ref{rome},\ref{max},\ref{inaf},\ref{iiass}} }

\institute{
Astrobiology Research Unit, Universit\'e de Li\`ege, All\'ee du 6 Ao\^ut 19C, B-4000 Li\`ege, Belgium \label{aru_liege}
\and Instituto de Astrof\'isica de Andaluc\'ia (IAA-CSIC), Glorieta de la Astronom\'ia s/n, 18008 Granada, Spain \label{iaa}
\and Space Sciences, Technologies and Astrophysics Research (STAR) Institute, Universit\'e de Li\`ege, All\'ee du 6 Ao\^ut 19C, B-4000 Li\`ege, Belgium \label{star_liege}
\and Cavendish Laboratory, JJ Thomson Avenue, Cambridge, CB3 0HE, UK \label{cavendish}
\and Instituto de Astrof\'{i}sica de Canarias (IAC), 38205 La Laguna, Tenerife, Spain \label{iac}
\and AIM, CEA, CNRS, Universit\'e Paris-Saclay, Universit\'e de Paris, F-91191 Gif-sur-Yvette, France \label{cea}
\and Kavli Institute for Astrophysics and Space Research, Massachusetts Institute of Technology, Cambridge, MA 02139, USA \label{mitkavli}
\and Department of Physics and Astronomy, University College London, Gower Street, London, WC1E 6BT, United Kingdom \label{clondon}
\and Departamento de Astrof\'{i}sica, Universidad de La Laguna (ULL), 38206 La Laguna, Tenerife, Spain \label{la_laguna}
\and Department of Astronomy \& Astrophysics, University of Chicago, Chicago, IL 60637, USA \label{chicago}
\and Center for Astrophysics and Space Sciences, UC San Diego, UCSD Mail Code 0424, 9500 Gilman Drive, La Jolla, CA 92093-0424, USA \label{san_diego}
\and Center for Interdisciplinary Exploration and Research in Astrophysics (CIERA), Northwestern University, 1800 Sherman, Evanston, IL, 60201, USA \label{North}
\and Department of Astrophysical \& Planetary Sciences, University of Colorado Boulder, 2000 Colorado Avenue, Boulder, CO 80309, USA \label{boulder}
\and Department of Earth and Planetary Sciences, University of California, Riverside, CA 92521, USA \label{riverside}
\and Department of Earth, Atmospheric and Planetary Science, Massachusetts Institute of Technology, 77 Massachusetts Avenue, Cambridge, MA 02139, USA \label{miteaps}
\and Department of Physics and Kavli Institute for Astrophysics and Space Research, Massachusetts Institute of Technology, Cambridge, MA 02139, USA \label{mit2}
\and Department of Physics \& Astronomy, Vanderbilt University, 6301 Stevenson Center Ln., Nashville, TN 37235, USA \label{vanderbilt}
\and Universit\'e de Montr\'eal, D\'epartement de Physique, IREX, Montr\'eal, QC H3C 3J7, Canada \label{montreal}
\and Observatoire du Mont-M\'egantic, Universit\'e de Montr\'eal, Montr\'eal H3C 3J7, Canada \label{montreal2}
\and NASA Exoplanet Science Institute, Caltech/IPAC, Mail Code 100-22, 1200 E. California Blvd., Pasadena, CA 91125, USA \label{nasa_exo}
\and NASA Ames Research Center, Moffett Field, CA 94035, USA \label{nasa_ames}
\and NASA Goddard Space Flight Center, 8800 Greenbelt Rd, Greenbelt, MD 20771, USA \label{nasa_goddard}
\and Department of Astronomy, University of Maryland, College Park, MD  20742, USA \label{maryland}
\and School of Physics \& Astronomy, University of Birmingham, Edgbaston, Birmingham B15 2TT, UK \label{birmingham}
\and Center for Astrophysics | Harvard \& Smithsonian, 60 Garden Street, Cambridge, MA, 02138, USA \label{cfa}
\and Center for Space and Habitability, University of Bern, Gesellschaftsstrasse 6, 3012, Bern, Switzerland \label{unibe}
\and Department of Physics, University of Warwick, Gibbet Hill Road, Coventry CV4 7AL, United Kingdom \label{warwick}
\and Physikalisches Institut, University of Bern, Gesellsschaftstrasse 6, 3012 Bern, Switzerland \label{unibe2}
\and Observatoire de Genève, Université de Genève, Chemin de Pegasi, 51, 1290 Versoix, Switzerland \label{obsgene}
\and Universidad Nacional Aut\'onoma de M\'exico, Instituto de Astronom\'ia, AP 70-264, Ciudad de M\'exico,  04510, M\'exico \label{ciudad}   
\and Universidad Nacional Aut\'onoma de M\'exico, Instituto de Astronom\'ia, AP 106, Ensenada 22800, BC, M\'exico \label{uname} 
\and European Space Agency (ESA), European Space Research and Technology Centre (ESTEC), Keplerlaan 1, 2201 AZ Noordwijk, Netherlands \label{estec}
\and Dpt. Física Teórica y del Cosmos, Universidad de Granada, Campus de Fuentenueva s/n, 18071 Granada, Spain \label{ugr} 
\and Oukaimeden Observatory, High Energy Physics and Astrophysics Laboratory, Faculty of sciences Semlalia, Cadi Ayyad University, Marrakech, Morocco \label{tn}
\and The CHARA Array, Georgia State University, USA \label{chara}
\and Steward Observatory and Department of Astronomy, The University of Arizona, Tucson, AZ 85721, USA \label{arizona}
\and Lunar and Planetary Laboratory and Department of Planetary Sciences, The University of Arizona, Tucson, AZ 85721, USA\label{arizona2}
\and Vatican Observatory, 00120 Citt\`a del Vaticano \label{vatican}
\and Graduate Institute of Astronomy, National Central University, Taoyuan 32001, Taiwan \label{taiwan}
\and Max-Planck-Institut f\"{u}r Astronomie, K\"onigstuhl 17, 69117 Heidelberg, Germany \label{max}
\and Department of Physics, University of Rome ``Tor Vergata’’, Via della Ricerca Scientifica 1, I-00133, Rome, Italy \label{rome}
\and INAF — Turin Astrophysical Observatory, via Osservatorio 20, I-10025, Pino Torinese, Italy \label{inaf}
\and International Institute for Advanced Scientific Studies (IIASS), Via G. Pellegrino 19, I-84019, Vietri sul Mare (SA), Italy \label{iiass}
}

\date{Received ...; accepted ...}

\abstract
 {Several planetary formation models have been proposed to explain the observed abundance and variety of compositions of super-Earths and mini-Neptunes. In this context, multitransiting systems orbiting low-mass stars whose planets are close to the radius valley are benchmark systems, which help to elucidate which formation model dominates.}
{We report the discovery, validation, and initial characterization of one such system, TOI-2096 (TIC 142748283), a two-planet system composed of a super-Earth and a mini-Neptune hosted by a mid-type M dwarf located 48~pc away.}  
{We characterized the host star by combining optical spectra, analyzing its broadband spectral energy distribution, and using evolutionary models for low-mass stars. Then, we derived the planetary properties by modeling the photometric data from TESS and ground-based facilities. In addition, we used archival data, high-resolution imaging, and statistical validation to support our planetary interpretation.}   
{We found that the stellar properties of TOI-2096 correspond to a dwarf star of spectral type M4$\pm$0.5. It harbors a super-Earth (R=1.24$\pm{0.07}$~R$_{\oplus}$) and a mini-Neptune (R=1.90$\pm{0.09}$~R$_{\oplus}$) in likely slightly eccentric orbits with orbital periods of 3.12 d and 6.39~d, respectively. These orbital periods are close to the first-order 2:1 mean-motion resonance (MMR), a configuration that may lead to measurable transit timing variations (TTVs). We computed the expected TTVs amplitude for each planet and found that they might be measurable with high-precision photometry delivering mid-transit times with accuracies of $\lesssim$2~min. Moreover, we conclude that measuring the planetary masses via radial velocities (RVs) could also be possible. Lastly, we found that these planets are among the best in their class to conduct atmospheric studies using the NIRSpec/Prism onboard the James Webb Space Telescope (JWST).}
{The properties of this system make it a suitable candidate for further studies, particularly for mass determination using RVs and/or TTVs, decreasing the scarcity of systems that can be used to test planetary formation models around low-mass stars.}


\maketitle

\section{Introduction}\label{sec:int}

The discovery of a large abundance of small transiting exoplanets with sizes ranging from slightly larger than the Earth to 4~R$_{\oplus}$ was one of the most relevant results of NASA's Kepler mission \citep{howard2012,fressin2013}. It is worth mentioning that the first exoplanet discovered with its size in this range was found by the CoRoT mission \citep{baglin2006}, named CoRoT-7b \citep{leger2009}, which was also the first small exoplanet with both measured radius and mass and, consequently, the first constraint on the bulk composition of a planet of this class \citep{queloz2009}. With additional discoveries by K2 and TESS, these worlds have reached >3000 out of the >5000 confirmed planets that exist to date.\footnote{Based on the NASA Exoplanet Archive in October 2022; \url{https://exoplanetarchive.ipac.caltech.edu/}}.

This category of planets has become the most abundant of the known planets in our Galaxy, seeming to exist around roughly 30--50$\%$ of all main-sequence stars \citep[see, e.g.,][and references therein]{raymond2022}. As these planets are not present in our Solar System, their abundance has challenged past planetary formation models \citep{ida2004,mordasini2009}, and the mechanisms that form them are still hotly debated  
\citep{howard2010,chiang2013,hansen2013,lee2014,ginzburg2016,adams2020,raymond2018,raymond2020,izidoro2021}. 
These planets also show intriguing physical properties. For example, combining Doppler surveys with photometry observations provided by ground- and space-based missions has revealed a diversity of compositions \citep{hatzes2015,LuquePalle2022}. Some are high-density "super-Earths", massive rocky planets \cite[$\rho$ $\gtrsim$ 5.0 g/cm$^3$, e.g.,][]{kreidberg2019,jindal2020,essack2023}. Others named "mini-Neptunes" have densities low enough to be consistent with the presence of extended H/He envelopes around a solid core \cite[$\rho$ $\lesssim$ 4 g/cm$^3$, e.g.,][]{leslie2011,kite2020}. Some planets in this radius range also appear to be water-worlds, that is, planets with water-mass fractions $\gtrsim$10$\%$ and lacking H/He atmospheres \citep{bean2021,adam2008,cadieux2022}. 

A key result of the California-Kepler Survey \citep{california2017a,california2017b} has shown that super-Earths and mini-Neptunes with orbital periods less than 100~d follow a bimodal size distribution with peaks at $\sim$1.3 and $\sim$2.7~R$_{\oplus}$ \citep{fulton2017}, defining the so-called "radius valley". Various mechanisms have been proposed to explain the paucity of planets in this particular range of radii, encompassing the transition between rocky super-Earths and gas-rich mini-Neptunes. Currently, two classes of formation scenarios have been proposed that yield opposing predictions. The first proposes that planets are born in a gas-poor inner protoplanetary disk and lack initial gaseous envelopes \citep{lee2014,lee2016,lee2021}. The second hypothesis proposes that planets are formed with H/He envelopes but experience atmosphere sculpting driven by either photoevaporation by the host star \citep{owen2013,lopez2013} or core-powered mass loss driven by heat left over from the planet's formation \citep{ginzburg2016,ginzburg2018,gupta2019}. If the gas-poor formation scenario is at work, the radius valley position is not affected by the stellar mass; if atmospheric erosion is at work, the radius valley depends on the stellar type, shifting to smaller planets for low-mass stars \citep{wu2019,berger2020,cloutier2020RV}. 

On the contrary, a recent study indicates that there is not a radius valley for low-mass stars but a "density gap" between super-Earths and mini-Neptunes at 0.65$\rho_{\oplus}$, with no overlap between these populations \citep{LuquePalle2022}. Using a sample of 34 exoplanets with a precise radius (to at least <8$\%$) and mass (to at least <25\%), the authors found evidence that the planetary radius does not have any dependence on the orbital period. Instead, they identify three different density regimes: rocky planets ($\rho=0.94\pm0.13~\rho_{\oplus}$), water worlds ($\rho=0.47\pm0.05~\rho_{\oplus}$), and puffy mini-Neptunes ($\rho=0.24\pm0.04~\rho_{\oplus}$). This result favors pebble accretion models as the main mechanism for forming small planets orbiting low-mass stars \citep{venturini2020,brugger2020}. In the case of puffy planets, they found a larger radii dispersion that could be a consequence of individual H/He accretion histories instead of atmospheric loss processes. This would imply that water worlds and puffy mini-Neptunes are part of the same population of planets. Hence, this result would impact our understanding of planetary formation if it is further confirmed with a larger sample of exoplanets. The authors attempted to extend their analysis to FGK stars; however, the low number of precisely characterized small planets around these stellar types hindered any conclusion.  

Then, the scientific community is still missing a robust planetary formation model that explains the observed population of super-Earths and mini-Neptunes. To this end, it is necessary to study the position of the radius valley and the density gap at different stellar types with a statistically robust sample of planets whose radii and masses enable estimates of their bulk densities and the mass fractions of their envelopes. In this context, multitransiting systems (those with more than one transiting planet) with planets close to the radius valley or density gap are of particular interest, as they enable direct comparative planetology, having evolved around the same host star and formed within the same protoplanetary disk \citep{kubyshkina2019,cloutier2020RV,owen2020}. While all stellar types are essential to test different formation mechanisms, low-mass stars (mid-K to mid-M dwarfs) are preferred targets. Indeed, their small sizes ease the discovery and characterization of small planets using transit and radial velocity (RVs) methods. These stars are also the most abundant in the solar neighborhood \citep{reyle2021}, and it has been found that, on average, they host a few small planets per star \citep[see e.g., ][]{clanton2014,dressing2015,ballard2016,tuomi2019}. Lastly, these stars emit X-rays and extreme ultraviolet radiation (XUV) for a more extended period \citep[up to 2~Gyr;][]{johnstone2015} than solar-type stars \citep[about 300~Myr;][]{gudel2004}, which increases the likelihood of atmospheric loss processes.  

Unfortunately, among the large population of small transiting exoplanets currently known, few are found in multitransiting systems orbiting low-mass stars. Hence, each new system in this category represents an excellent opportunity to increase our understanding of how planetary systems form and evolve.

Here we present the discovery, validation, and characterization of a multiplanetary system with at least two transiting planets orbiting the nearby mid-type M dwarf TOI-2096. The two planets have sizes of 1.2~R$_{\oplus}$ and 1.9~R$_{\oplus}$, flanking both sides of the radius valley. In addition, the orbital periods of the planets are close to the 2:1 mean motion resonance (MMR), allowing the measurement of planetary masses using both RVs and transit timing variations (TTVs) methods. Furthermore, its two planets are also promising targets for a detailed atmospheric characterization with the James Webb Space Telescope (JWST). All of these properties make this system a compelling case for further investigation.

The paper is structured as follows. 
In Sect.~\ref{sec:obs}, we present the facilities and data used for validating and characterizing the system. 
In Sect.~\ref{sec:star}, we provide all of the available information on the host star and its characterization. 
In Sect.~\ref{sec:validation}, we validate the planetary nature of the two signals. 
In Sect.~\ref{sec:ta}, we describe the analyses that allow us to derive the planetary parameters and search for TTVs. 
In Sect.~\ref{sec:search}, we search for additional transiting planets and establish detection limits from our data set. 
In Sect.~\ref{sec:dyn}, we conduct a dynamical analysis of the system. 
In Sect.~\ref{sec:disc}, we discuss in detail several aspects of our results, including prospects for RV follow-up to measure the planetary masses, the planetary sizes in the context of the radius valley, and future atmospheric characterization. 
Finally, in Sect.~\ref{sec:concl}, we present our conclusions.

\section{Observations}\label{sec:obs}
This section details all of the observations and facilities used to study the TOI-2096 system. This effort corresponds to time-series photometry measured in six TESS sectors and in which the planets were originally found, plus observations gathered by seven ground-based telescopes used for the photometric follow-up, and by two optical spectrographs and a high-resolution imager to accurately characterize the star.

\begin{table*}
 \begin{center}
 \begin{tabular}{l l c c c c}
 \toprule
Candidate & Date (UT) & Telescope & Bandpass & Exp. time (s) & $\Delta$ln$Z$ \\
 \midrule
 \multirow{16}{*}{TOI-2096\,b} & 07 Sep 2020 & OMM-1.6m & $i'$ & 30 & 3.9$^{\dag}$ $^{\ddagger}$\\
  & 01 Oct 2020 &  TRAPPIST-North-0.6m & $I+z'$ & 80 & 4.0$^{\dag}$ $^{\ddagger}$ \\
  & 26 Oct 2020 &  TRAPPIST-North-0.6m & $I+z'$ & 80 & <0 \\
  & 23 Nov 2020 &  TRAPPIST-North-0.6m & $I+z'$  & 80 & 3.2$^{\dag}$ $^{\ddagger}$ \\
  & 29 Nov 2020 & SAINT-EX-1.0m & $z'$ & 60 & <0 \\
  & 02 Dec 2020 & SAINT-EX-1.0m & $z'$ & 60 & 7.0$^{\dag}$ $^{\ddagger}$\\
  & 18 Dec 2020 & Artemis-1.0m & $I+z'$ & 30 & <0  \\ 
  & 21 Dec 2020 & Liverpool-2.0m & $SDSS-Z$ &  70 & 6.1$^{\dag}$ $^{\ddagger}$\\
  & 21 Dec 2020 & Artemis-1.0m & $z'$ & 30 & 5.7$^{\dag}$ \\
  & 24 Dec 2020 & LCO-1.0m & $Ic$ & 90 & <0 \\
  & 15 Jan 2021 & TRAPPIST-North-0.6m & $I+z'$ & 80 & 4.4$^{\dag}$ \\
  & 09 Feb 2021 & Artemis-1.0m & $I+z'$  & 30 & 7.0$^{\dag}$ $^{\ddagger}$\\
  & 12 Feb 2021 & Artemis-1.0m & $I+z'$ & 30 & 3.7$^{\dag}$ $^{\ddagger}$\\
  & 03 Mar 2021 & TRAPPIST-North-0.6m & $I+z'$  & 80 & 5.1$^{\dag}$ $^{\ddagger}$\\
  & 03 Mar 2021 & Artemis-1.0m & $I+z'$ & 30 & 1.6 \\
  & 09 Mar 2021 & Artemis-1.0m & $I+z'$  & 30 & 1.2 \\
  & 04 May 2021 & VATT-1.83m & $GG495$ & 45 & 2.1 \\
\midrule
 \multirow{7}{*}{TOI-2096\,c} & 11 Oct 2020 & TRAPPIST-North-0.6m & $I+z'$ & 80 & 5.4$^{\dag}$ $^{\ddagger}$\\
  & 14 Dec 2020 & Artemis-1.0m & $I+z'$ & 30 & 1.5 \\ 
  & 20 Dec 2020 & LCO-1.0m & $Ic$ & 90 & 7.7$^{\dag}$ $^{\ddagger}$\\
  & 27 Dec 2020 & TRAPPIST-North-0.6m & $I+z'$ & 80 & 0.8 \\
  & 15 Jan 2021 & TRAPPIST-North-0.6m & $I+z'$ & 80 & 4.4$^{\dag}$ $^{\ddagger}$\\
  & 16 Feb 2021 & Artemis-1.0m & $I+z'$ & 30 & 12.8$^{\dag}$ $^{\ddagger}$\\
  & 20 Mar 2021 & Artemis-1.0m & $I+z'$ & 30 & 7.5$^{\dag}$ $^{\ddagger}$\\

 \hline 
 \end{tabular}
 \caption{Ground-based time-series photometric observations of TOI-2096. Only those with Bayes factor $\Delta$ ln Z>2.3 (highlighted with a $\dag$ symbol) are used for the global analysis.
 Those marked with a $\ddagger$ symbol were used for the analyses of TTVs (see Sect.~\ref{sec:ta} for details).}
 \label{tab:GBobservations}
 \end{center}
 \end{table*}

\subsection{TESS photometry}\label{subsec:tess}

TOI-2096 (TIC~142748283) is a low-mass star included in the TESS Candidate Target List, a ranked list from the TESS Input Catalog (TIC) prioritizing stars where it would be easier to detect small transiting planets \citep{stassun2018}. TOI-2096 was observed by TESS with 2-min-cadence in Sectors 14, 20, and 21 during the primary mission\footnote{Sector 14 (2019-Jul-18 to 2019-Aug-15, in cycle 2); Sector 20 (2019-Dec-24 to 2020-Jan-21, in cycle 2); Sector 21 (2020-Jan-21 to 2020-Feb-18, in cycle 2).}. During the extended mission, it was observed using the same cadence in Sectors 40, 41, 47, and 53, and it will be reobserved in Sector 60\footnote{Sector 40 (2021-Jun-24 to 2021-Jul-23, in cycle 4); Sector 41 (2021-Jul-23 to 2021-Aug-20, in cycle 4); Sector 47 (2021-Dec-30 to 2022-Jan-28, in cycle 4); Sector 53 (2022-Jun-13 to 2022-Jul-09, in cycle 4); Sector 60 (2022-Dec-23 to 2023-Jan-18, in cycle 5).}. After the combination of sectors observed during the primary mission, the TESS Science Processing Operations Center \citep[SPOC;][]{jenkins2016} conducted a transit search on 5 May 2020 with a noise-compensating matched filter \citep{jenkins2002,jenkins2010,jenkins2020}, yielding the detection of two signals: TOI-2096.01 and .02. The TESS Science Office issued alerts for these planetary candidates on 15 July 2020 \citep{guerrero2021}.  These two signals were strengthened with the additional sectors during the extended mission. The signals were found to have orbital periods of 3.12~d and 6.39~d and transit depths of 3.6 and 6.9~ppt for candidates .01 and .02, respectively. Hereafter, we refer to TOI-2096.01 as TOI-2096\,b, and TOI-2096.02 as TOI-2096\,c. In this study, we used the observations from Sectors 14, 20, 21, 40, 41, and 47. Sectors 53 and 60 will be included in a forthcoming study focused on planetary mass determination through TTVs using high-precision photometry. 

To perform our analyses, we retrieved the Presearch Data Conditioning Simple Aperture Photometry (PDCSAP) fluxes, which have been corrected for instrument systematics and for crowding effects \citep{stumpe2012,smith2012,stumpe2014} from the Mikulski Archive for Space Telescopes. From each sector, we
removed the data points flagged as `bad quality', which correspond to 12$\%$, 12$\%$, 10$\%$, 6$\%$, 8$\%$, and 6$\%$ of the data in the six sectors, respectively.
 
\begin{figure*}
    \centering
    \includegraphics[width=0.999\textwidth]{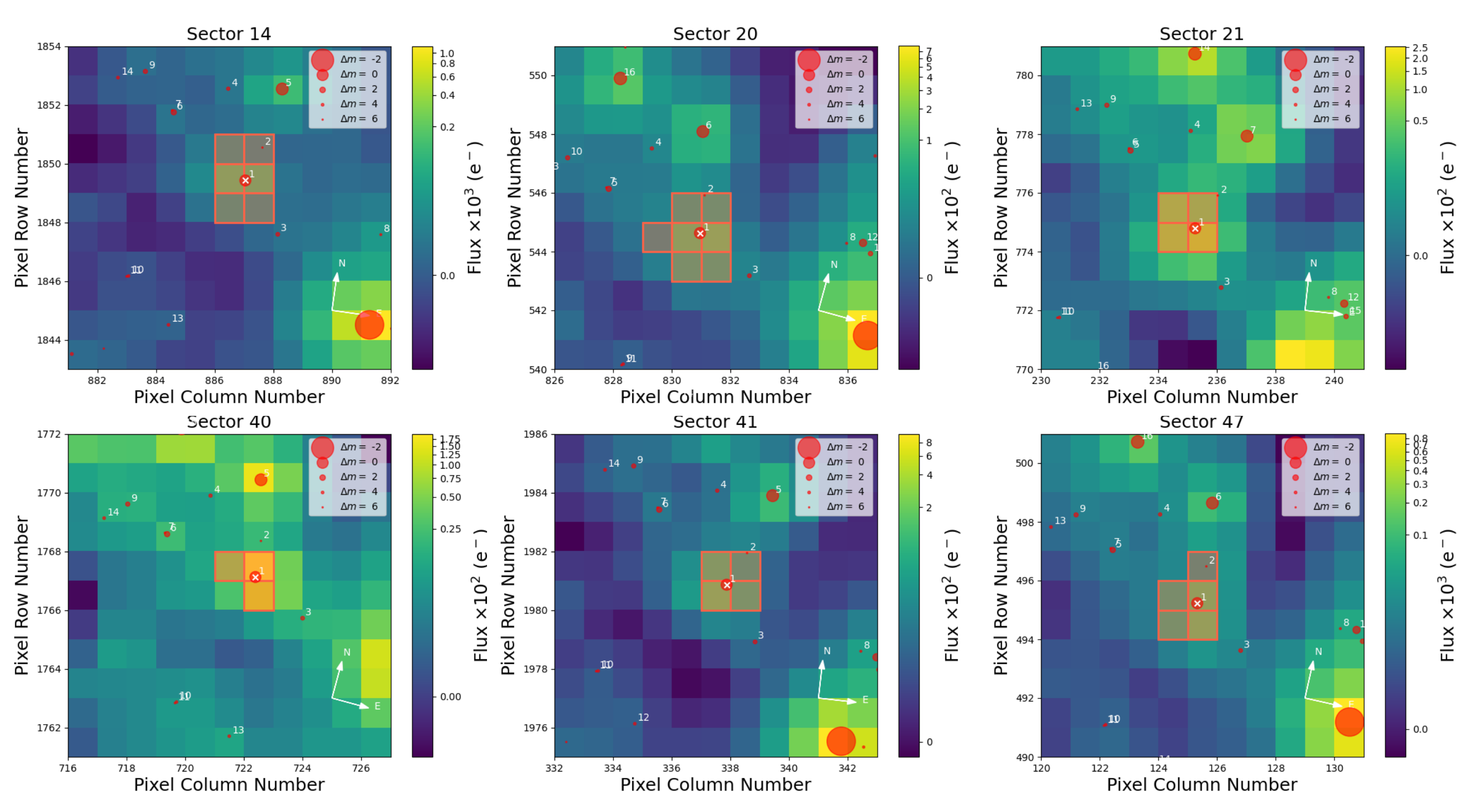}
    \caption{\textit{TESS} target pixel files (TPFs) of the six sectors that observed TOI-2096 generated by means of \texttt{tpfplotter} (\citealp{aller:2020}). The apertures used to extract the photometry by the SPOC pipeline are shown as red-shaded regions. The {\it Gaia\/} DR2 catalog is overplotted, with all sources up to six magnitudes in contrast with TOI-2096 shown as red circles. We note that the symbol size scales with the magnitude contrast.}
    \label{fig:tpf}
\end{figure*}

\begin{figure*}
    \centering
    \includegraphics[width=0.999\textwidth]{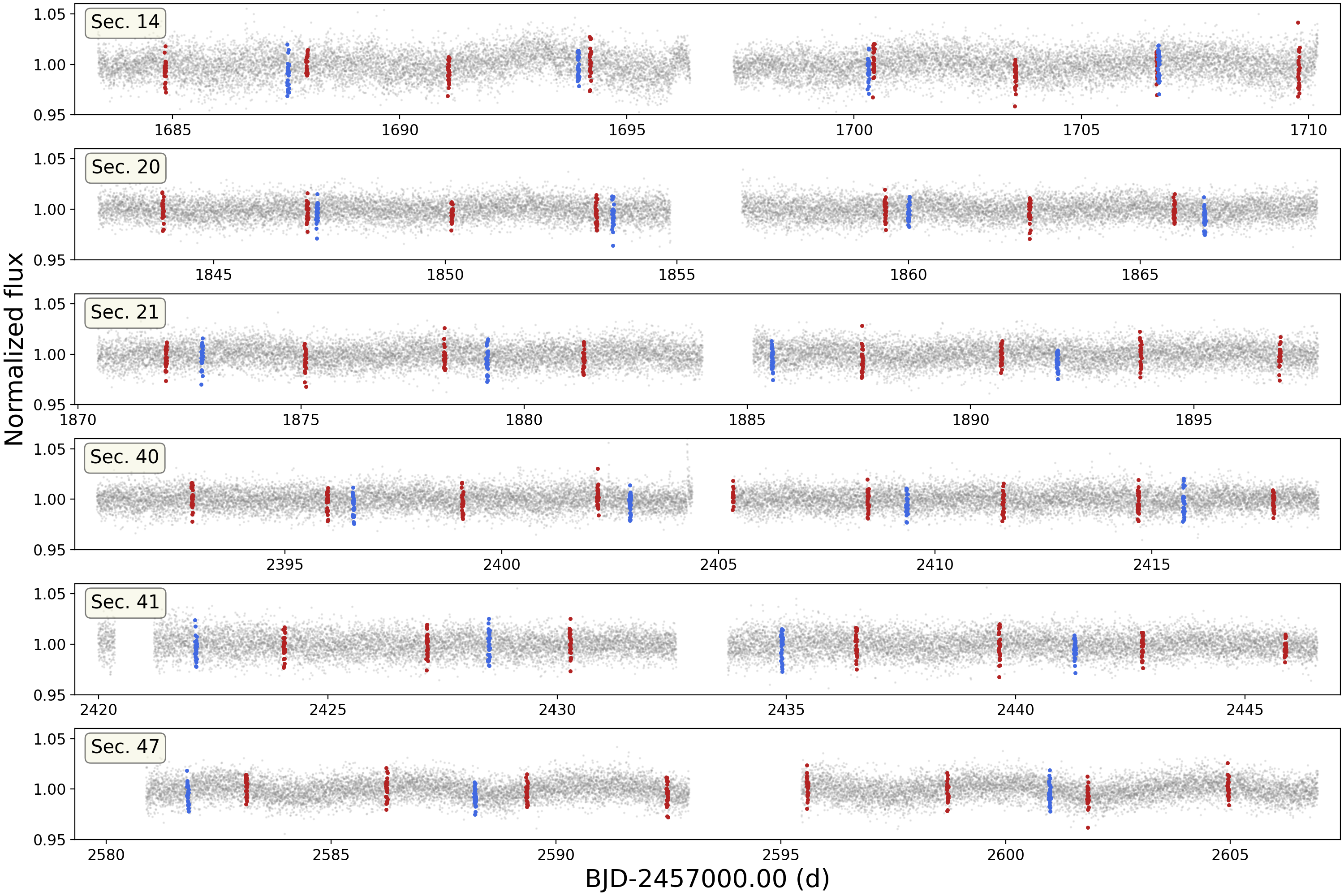}
    \caption{\textit{TESS} photometric time-series of TOI-2096, obtained for sectors 14, 20, 21, 40, and 47. In all cases, the gray points correspond to the PDCSAP fluxes obtained from the SPOC pipeline. The red and blue points correspond to the location of the transits for the candidates TOI-2096.01 and TOI-2096.02, respectively.}
    \label{fig:lc}
\end{figure*}

Fig.~\ref{fig:tpf} shows the TESS fields of view and the photometric apertures used over each of the six sectors examined in this study, with the location of nearby \textit{Gaia} DR2 \citep{gaia18} sources superimposed. The target star was well-isolated, with no bright stars within the aperture. The closest star\footnote{\textit{Gaia} DR2: 1126422935075485184} is a considerably fainter source ($\Delta$m$\sim$5), one pixel away and not always included in the aperture. In addition, Fig.~\ref{fig:lc} displays the time-series photometry for each of these sectors, highlighting the location of the transits corresponding to the two signals alerted by SPOC.

\subsection{Ground-based follow-up photometry}
\label{sec:gbobservations}

We used ground-based follow-up photometry from several facilities in the context of the TESS Follow-up Observing Program (TFOP) Sub-Group 1 (SG1) for Seeing-Limited Photometry. The main goals of these observations were to confirm the signals corresponding to TOI-2096\,b and c in the target star at transit times, obtain light curves with higher precision than TESS data, and evaluate the transit depths in different bands to assess the chromatic dependence. 

To obtain a dataset as homogeneously as possible, we extracted the photometry from all the ground-based observations using the \texttt{prose} Python package\footnote{\href{https://github.com/lgrcia/prose}{https://github.com/lgrcia/prose}} \citep{prose}. Much like AstroImageJ \citep[AIJ;][]{ collins2017}, \texttt{prose} is an instrument-agnostic framework to reduce raw FITS images. Its architecture allows building fully custom pipelines using a wide range of processing blocks while providing preimplemented options for common processing tasks, one of which is a basic photometric extraction pipeline. 

In the following subsubsections, we describe all the ground-based observations, which are also summarized in Table~\ref{tab:GBobservations}.
With the exception of SAINT-EX and Artemis, all observations were scheduled using the TESS Transit Finder (TTF) tool, a customized version of the \texttt{Tapir} software \citep{jensen2013}.

\subsubsection{TRAPPIST-North}
We observed five full transits of TOI-2096\,b and three full transits of TOI-2096\,c using the 0.6-m TRAPPIST-North telescope, located at Oukaïmeden observatory, Morocco \citep{jehin2011,gillon2011}. TRAPPIST-North is equipped with a 2048$\times$2048 Andor IKON-L BEX2-DD camera, providing a field of view of 20$^{\prime}$ $\times$ 20$^{\prime}$, and a pixel scale of 0.60$^{\prime\prime}$ per pixel. The telescope has a Ritchey-Chretien optical design with F/8 and a German equatorial mount, leading to a meridian flip in some observations. 
All the light curves were obtained using the $I+z'$ band and individual frame exposure times of 80~s. 

\subsubsection{SAINT-EX and Artemis}

SAINT-EX and Artemis are 1-m telescopes located at the Observatorio Astron\'{o}mico Nacional in San Pedro M\'{a}rtir (Mexico), and at the Teide Observatory in the Canary Islands (Spain), respectively. Both telescopes are part of the SPECULOOS network of six identical robotic telescopes \citep{delrez2018,sebastian2020} devoted to the search for terrestrial exoplanets orbiting the nearest ultracool dwarfs \citep{burdanov2018,gillon2018}. The facilities are Ritchey-Chretien telescopes equipped with ANDOR iKon-L BEX2-DD cameras and 2048$\times$2088 e2v CCD detectors, with a field of view of 12$^{\prime}$ $\times$ 12$^{\prime}$ and a pixel scale of 0.35$^{\prime\prime}$ per pixel. Using the Artemis telescope \citep{burdanov2022}, we observed six full transits of TOI-2096\,b and three of TOI-2096\,c, using $I+z'$ and $z'$ filters and exposure times of 30~s. Using SAINT-EX, we observed two full transits of TOI-2096\,b using the $z'$ band and an exposure time of 60~s (see Table~\ref{tab:GBobservations}). These observations were scheduled using the \texttt{SPOC} package, a dedicated planning tool for the SPECULOOS project \citep{sebastian2020}.

\subsubsection{OMM}
We observed one full transit of TOI-2096\,b with the PESTO 1024$\times$1024 pixel EM CCD camera installed on the 1.6-m telescope at Observatoire du Mont-Mégantic (OMM), Canada. PESTO has an image scale of 0.466$^{\prime\prime}$ per pixel, providing an on-sky 7.95$^{\prime}$ $\times$ 7.95$^{\prime}$ field of view. The observations were taken in the $i'$ filter with a 30-s exposure time.

\subsubsection{Las Cumbres Observatory}

Full transits of TOI-2096\,b and c were each observed once from the Las Cumbres Observatory McDonald site via the Las Cumbres Observatory Global Telescope (LCOGT) network \citep{Brown2013}. These observations were done with a 1-m telescope equipped with a \textit{Sinistro} camera, which has a plate scale of 0.389$^{\prime\prime}$ and a field of view of 26.4$^{\prime}$ $\times$ 26.4$^{\prime}$. The observations were taken using the $Ic$ band with 90-s exposures. The observations were partially defocused to smear the point-spread function (PSF) over more pixels, reducing error from uncertainties in the flat field.

\subsubsection{Liverpool}

We observed one full transit of TOI-2096\,b using the robotic 2-m Liverpool telescope, located at the Observatorio del Roque de los Muchachos, on the Canary island of La Palma, Spain. The observations were carried out with the 2048 $\times$ 2056 pixel optical wide field camera (2$\times$2 binned) equipped with an e2V CCD 231-84 detector, yielding a field of view of 10$^{\prime}$ $\times$ 10$^{\prime}$, and a pixel size of 0.30$^{\prime\prime}$ per pixel. We used the Sloan Digital Sky Survey (SDSS; \citealt{2000AJ....120.1579Y}) $Z$ band with a 70~s exposure time.

\subsubsection{VATT}

The Vatican Advanced Technology Telescope (VATT), or properly called the "Alice P. Lennon Telescope and the Thomas J. Bannan Astrophysics Facility” , is a 1.83-m telescope at the Mount Graham International Observatory (MGIO), in Arizona, USA. 
The VATT telescope is equipped with a 4096$\times$4096 pixel camera, providing a field of view of 12.5$^{\prime}$ $\times$12.5$^{\prime}$ and a pixel scale of 0.188$^{\prime\prime}$ per pixel. Our observations were executed as part of the Exo-Earth Discovery and Exploration Network (EDEN) program \citep[e.g.,][]{Gibbs2020}. We observed one full transit of TOI-2096\,b using the $GG495$ filter and an exposure time of 45~s.

\subsection{Spectroscopy}

To characterize the star TOI-2096, we used the Alhambra Faint Object Spectrograph and Camera (ALFOSC) mounted on the 2.5-m Nordic Optical Telescope (NOT) at the Observatorio Roque de los Muchachos (ORM) in La Palma, Canary Islands on 17 June 2021. The observing mode was long-slit spectroscopy with the gr5 grism, which covers the wavelength range 5000–10\,700 \AA. The slit width was 1.3\\arcsec, which resulted in a resolving power of $\sim$320. Two 120~s exposures of the target were taken. The spectrophotometric standard SP1446+259 was also observed to correct for the instrumental response. Some bias, continuum lamp, and arc-lamp images were also obtained to perform the bias and flat-field corrections and wavelength calibration.

\begin{figure}
    \centering
    \includegraphics[width=0.48\textwidth]{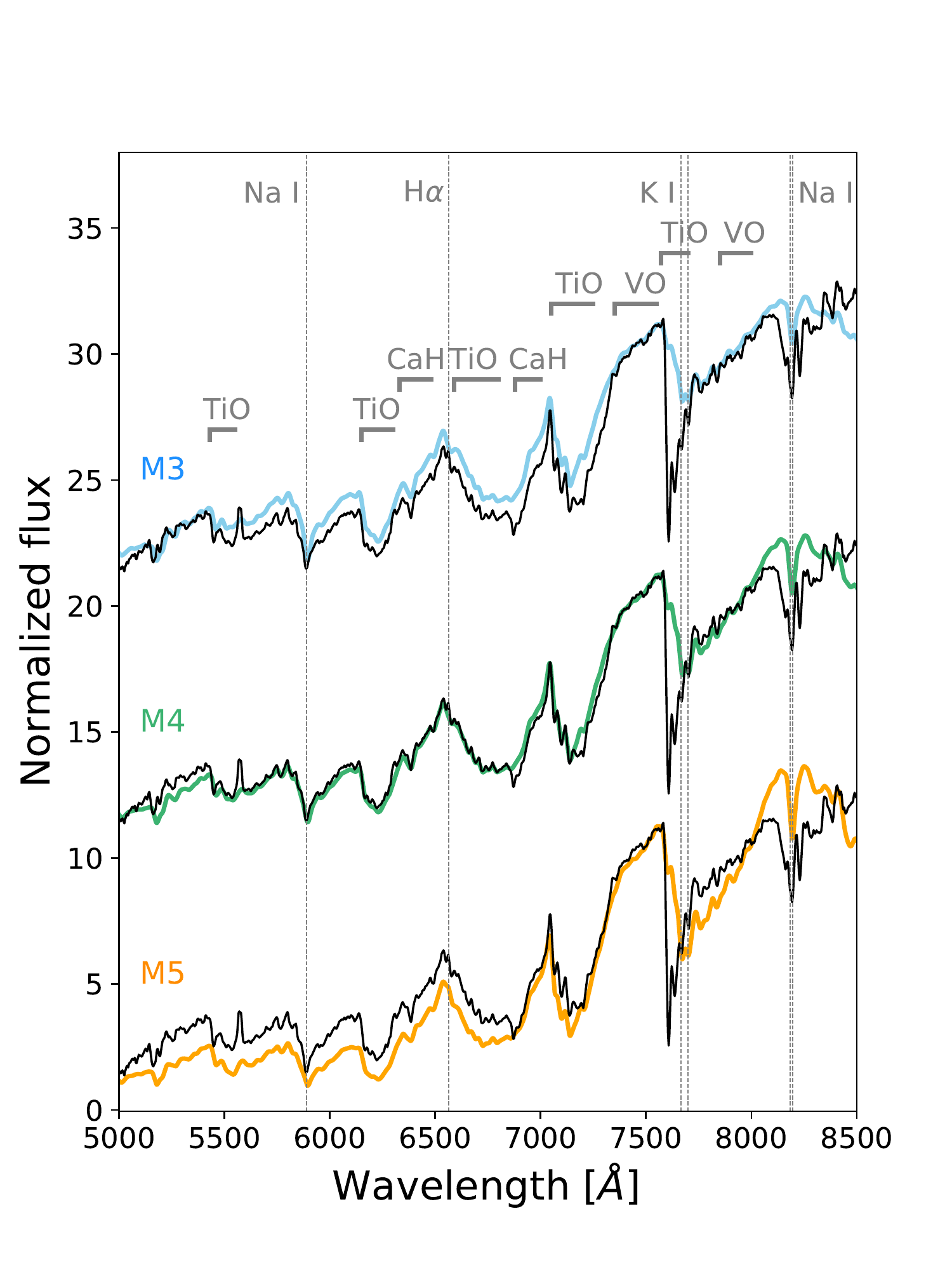}
    \caption{NOT/ALFOSC optical spectrum of TOI-2096 (black) compared to SDSS M3, M4, and M5 templates \citet{Bochanski2007} and \citet{kesseli:2017:16} (colored). The SDSS spectra were convolved using a Gaussian function to match the ALFOSC spectrum resolution. The spectra were normalized at $\sim$7500 \AA\, and shifted by a constant. Some remarkable absorption bands and lines are labeled in gray.}
    \label{fig:ALFOSCspT}
\end{figure}

The data were reduced using standard IRAF routines \citep{iraf1986}. The raw 2D spectral images were bias subtracted, flat-field corrected, and combined. Then, the spectra were extracted using the IRAF APALL routine, and the wavelength was calibrated using the thorium-argon arc-lamp images. Finally, the spectrum of the target was corrected from instrumental response using the spectrophotometric standard images. The spectrum was not corrected for atmospheric telluric absorption. The reduced spectrum is shown in Fig.~\ref{fig:ALFOSCspT}.

\begin{figure}
    \centering
    \includegraphics[width=\columnwidth]{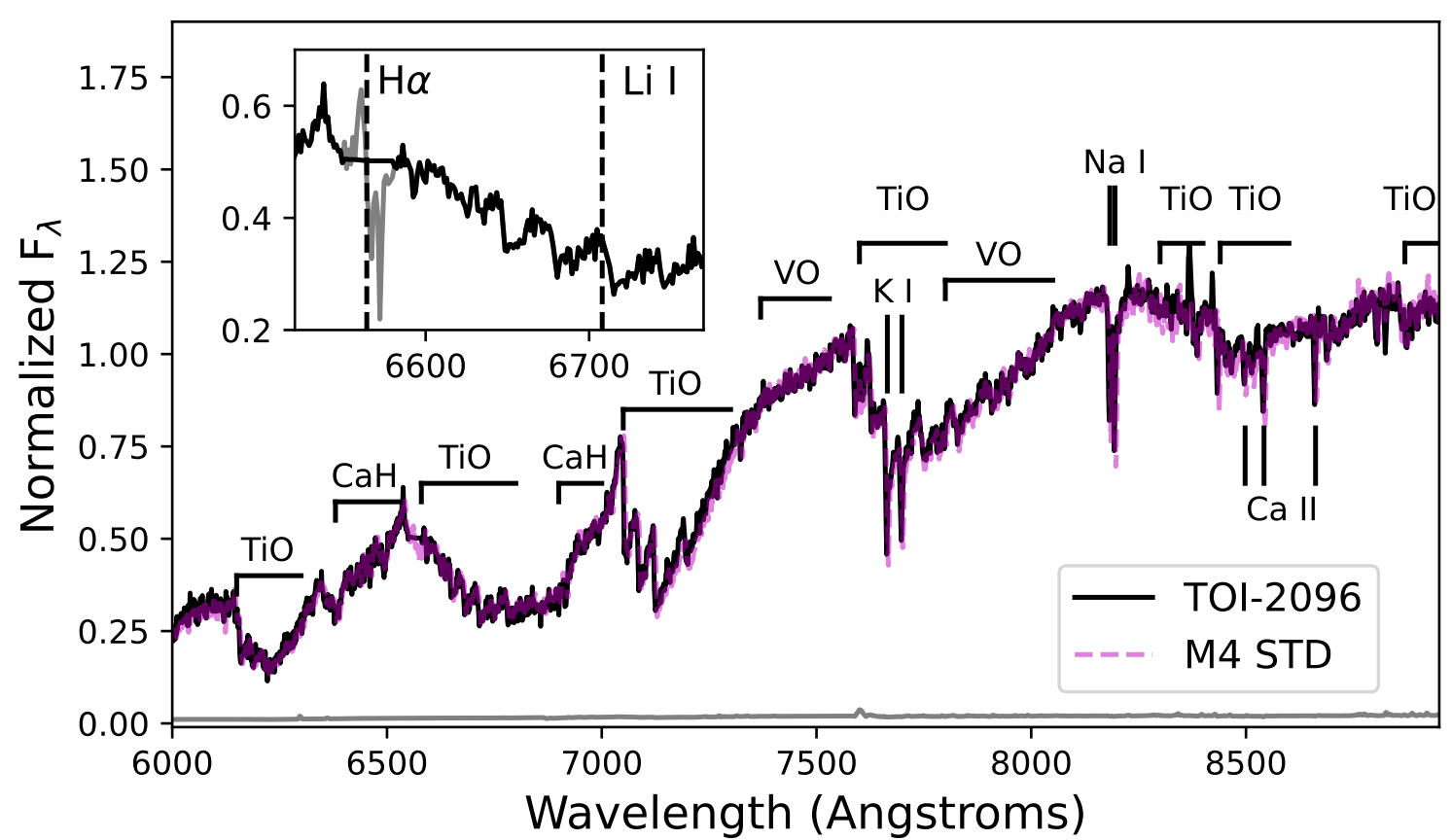}
    \caption{
        Shane/Kast red optical spectrum of TOI-2096 (black), compared to an M4 template \citep[magenta]{kesseli:2017:16}. Spectra are normalized in the 7400--7500\,{\AA} region, and major absorption features are labeled, including resolved lines. 
        The inset box shows a close-up of the region hosting H$\alpha$ and Li\,\textsc{i} features. The former is obscured by an unfortunate cosmic ray strike (unmasked spectral data in grey); the latter is not detected. 
    }
    \label{fig:kast}
\end{figure}

We obtained a slightly higher resolution optical spectrum of TOI-2096 with the Kast Double Spectrograph on the Lick Observatory 3-m Shane Telescope on 15 May 2021 (UT) in partly cloudy conditions with 2$\arcsec$ seeing. TOI-2096 was observed using the 2$\arcsec$ slit and 600/7500 red side grating, which provides a resolution of 1.3 \AA/pixel$^{-1}$ ($\lambda/\Delta\lambda$ $\approx$ 1900) over the wavelength range 6000--9000~\AA. Two exposures of 600~s each were obtained at an airmass of 1.27, and we observed the flux calibrator Feige 66 \citep{1992PASP..104..533H,1994PASP..106..566H} and the G2~V star HD~66171 that night with the same settings for flux and telluric calibration, respectively. 
Data were reduced using the KastRedux package,\footnote{\url{https://github.com/aburgasser/kastredux}.} which included image reduction (flat-fielding, bad pixel masking, and linearity correction) based on flat-field lamp exposures, source extraction, wavelength calibration using HeNeArHg arc lamp exposures (precision 0.56~{\AA} = 25 km/s), 
and flux calibration using the Feige 66 observation, and
telluric correction using the G2~V star observation.
The reduced spectrum is shown in Fig.~\ref{fig:kast} and has a median S/N of $\sim$40 at 7400~{\AA}.

\subsection{High-angular resolution imaging}
\label{sec:hri}

\begin{figure}
    \centering
    \includegraphics[width=0.48\textwidth]{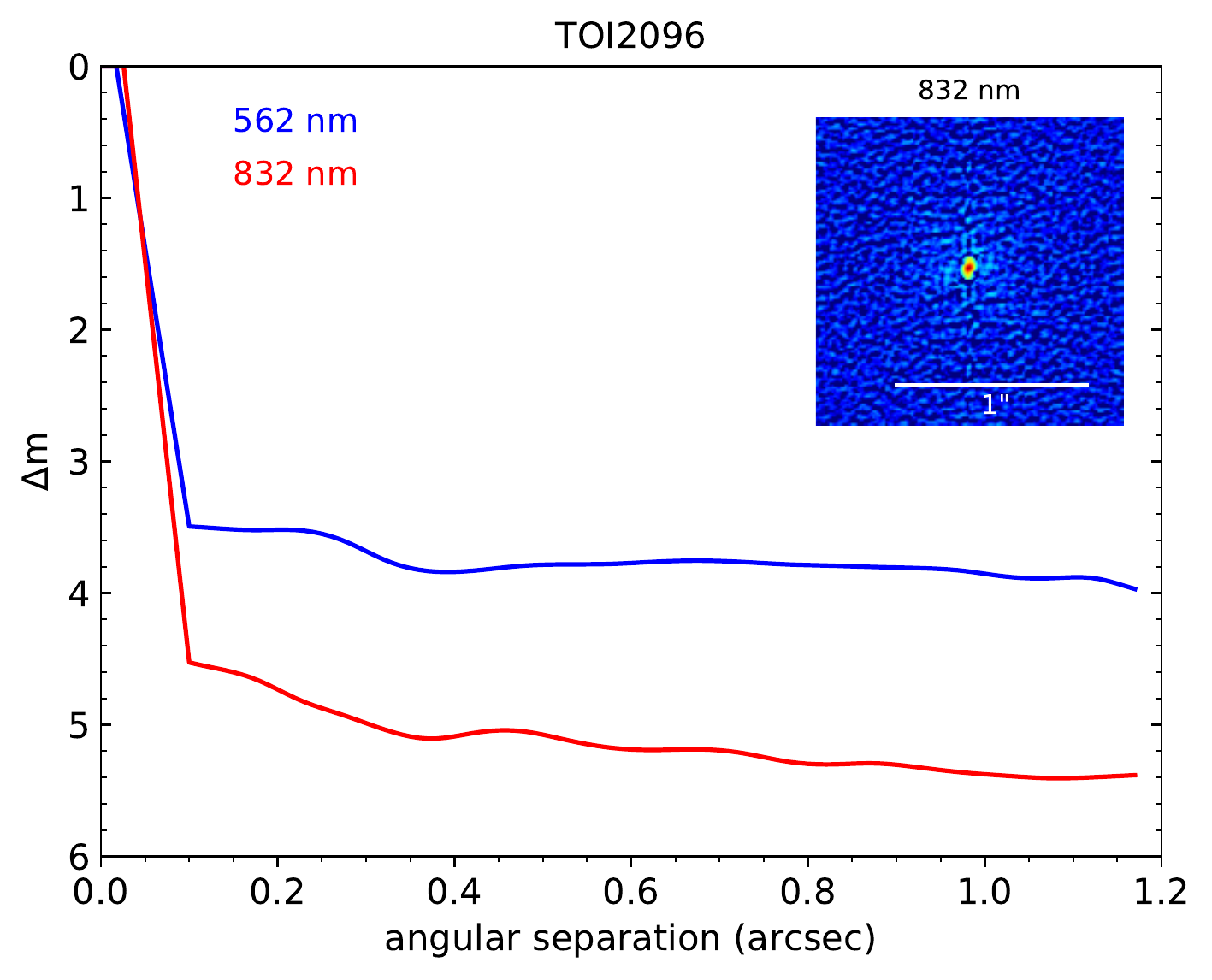}
    \caption{'Alopeke speckle imaging 5$\sigma$ contrast curves, along with the reconstructed 832-nm image.}
    \label{fig:hri}
\end{figure}

TOI-2096 was observed on 2021 February 02 UT using the 'Alopeke speckle instrument on Gemini North\footnote{https://www.gemini.edu/sciops/instruments/alopeke-zorro/}\citep{scott2021}. 'Alopeke provides simultaneous speckle imaging in two bands (562~nm and 832~nm) with output data products including a reconstructed image with robust contrast limits on companion detections \citep[e.g.,][]{howell2016}. Fifteen sets of 1,000 $\times$ 0.06~s exposures were collected and subjected to Fourier analysis in our standard reduction pipeline \citep[see][]{howell2011}. Fig.~\ref{fig:hri} shows our final contrast curves and the 832-nm reconstructed speckle image. We find that TOI-2096 is a single star with no companion brighter than $\Delta{m} \approx$ 4.5--5 magnitudes at 832 nm going from 0.1" to 1.2". From the diffraction limit ($\sim$0.02") to 0.1" the achieved contrasts are smaller, closer to 3 magnitudes, with no companion detected brighter than the $\Delta{m}$ contrast limits. At the distance of TOI-2096 ($d=48.5$\,pc) these angular limits correspond to spatial limits of 0.97~au to 58~au.

\section{Stellar characterization}\label{sec:star}

 In this section, we detail the methodology followed to determine the stellar properties of TOI-2096, summarized in Table~\ref{table:stellar}, and we provide the photometric and astrometric parameters obtained from the literature in Table~\ref{tab:starlit}.

\subsection{Spectroscopic analysis}


Both ALFOSC and Kast optical spectra were compared to optical spectral templates from \citet{Bochanski2007} and \citet{kesseli:2017:16}, derived from spectra obtained by the SDSS. For the ALFOSC spectrum, these templates were convolved with a Gaussian function to match the instrument's spectral resolution. For the Kast spectrum, the templates and spectrum have similar resolutions. Figs.~\ref{fig:ALFOSCspT} and~\ref{fig:kast} both show that the spectrum of TOI-2096 is well-matched to the M4 template, while Fig.~\ref{fig:ALFOSCspT} shows that the spectrum noticeably differs from M3 and M5 templates. We, therefore, assign a spectral type of M4$\pm$0.5 to this source. We also found the same classification using the index-based methods of \citet{2003AJ....125.1598L} and \citet{2007MNRAS.381.1067R}.

We evaluated the metallicity of TOI-2096 through the $\zeta$ index defined in \citet{2007ApJ...669.1235L}. From the Kast data, we find values of 0.91 $\leq \zeta \leq$ 0.97 using various calibrations \citep{2012AJ....143...67D,2013AJ....145..102L,2019ApJS..240...31Z}, all consistent with a dwarf metallicity class. The empirical metallicity/$\zeta$ calibration of \citet{2013AJ....145...52M} implies [Fe/H] = $-$0.03$\pm$0.20, that is to say a nearly solar metallicity star.

The optical data also cover the 6563~{\AA} H$\alpha$ line, commonly in emission in active mid-type M dwarfs \citep{1996AJ....112.2799H}. This line is undetected in the ALFOSC, while the Kast data had an unfortunate cosmic ray at the location of the H$\alpha$ line. We were also able to establish the absence of Li~I absorption at 6708~{\AA} in the Kast data, ruling out this object as a young brown dwarf \citep{1993ApJ...404L..17M}.


\subsection{Estimated age}


Several spectral indicators indicate that TOI-2096 is an average-aged field star. The absence of Li\,\textsc{i} absorption in the Kast optical spectrum implies an age $\gtrsim$30~Myr \citep{2004ApJ...604..272B}. Similarly, the absence of H$\alpha$ emission in the NOT/ALFOSC optical spectrum implies an age  $\gtrsim$4~Gyr \citep{west:2008:785}. We report this latter, more stringent constraint as our age estimate in Table\,\ref{table:stellar}.

\subsection{SED analysis, empirical relations, and evolutionary modeling}\label{sec:stellar_cha}

\begin{figure}
    \centering
    \includegraphics[width=\linewidth,trim=20 10 20 20,clip]{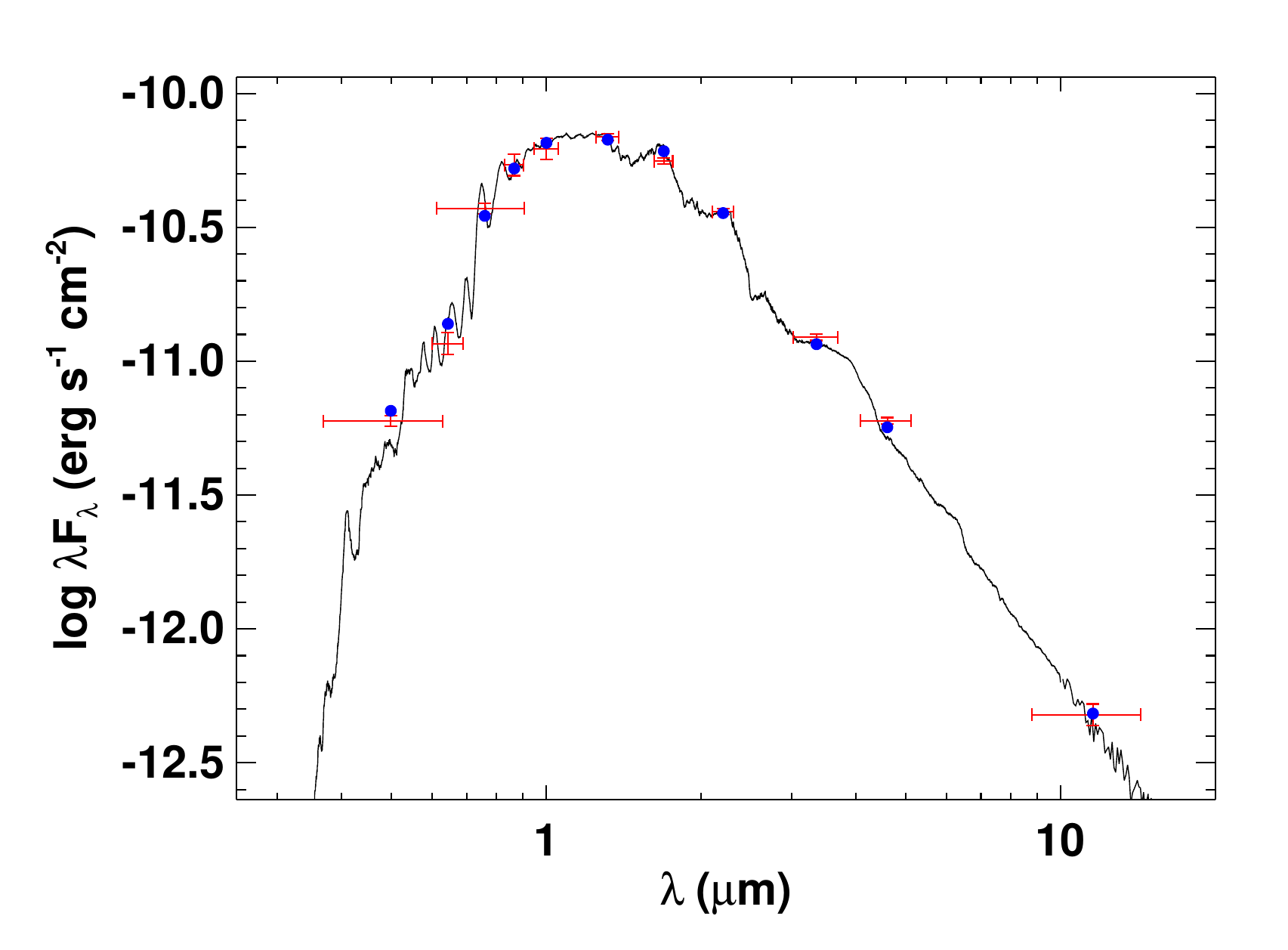}
    \caption{Spectral energy distribution (SED) fit of TOI-2096. The black curve is the best-fitting PHOENIX NextGen atmosphere model, red symbols are the observed fluxes (horizontal bars represent the effective bandpass widths), and blue symbols are the model fluxes.}
    \label{fig:sed}
\end{figure}

We analyzed the broadband SED to infer the basic stellar parameters following the approach of \citet{stassun:2016_SED} and \citet{stassun:2017_masses_radii, stassun:2018_masses_radii}.
In tandem with the \textit{Gaia} EDR3 \citep{gaia:2016_mission_paper, gaia:2021_EDR3} parallax \citep[with no systematic offset applied; see, e.g.,][]{stassun:2021_eDR3_offset}, we used the \textit{Pan-STARRS grizy}, \textit{2MASS JHK$_s$}, and \textit{WISE} W1--W3 magnitudes for the analysis. 
The empirical SED spans the 0.4--10\,$\mu$m wavelength range (Fig.~\ref{fig:sed}). 

We fit these data to the grid of PHOENIX dusty atmosphere models \citep[see][and references therein]{Allard:2012} to constrain on the stellar effective temperature ($T_\mathrm{eff}$), surface gravity ($\log g$), and metallicity ([Fe/H]).
Given the star's proximity, we fixed the extinction ($A_V$) to zero.
The resulting fit has a reduced $\chi^2$ of 1.4 and gives $T_\mathrm{eff} = 3300 \pm 75$\,K,
$\log g = 5.0 \pm 0.5$, and [Fe/H]$ = 0.0 \pm 0.3$, the last value consistent with the $\zeta$ analysis.
Directly integrating the (unreddened) model SED gives a bolometric flux at Earth of 
$F_\mathrm{bol} = 8.066 \pm 0.093 \times 10^{-11}$\,erg\,s$^{-1}$\,cm$^{-2}$.
Together with the \textit{Gaia} eDR3 parallax, $F_\mathrm{bol}$ and $T_\mathrm{eff}$ imply a stellar radius of $R_\star = 0.235 \pm 0.011\,R_{\sun}$. 

We also inferred constraints on the stellar parameters using the evolutionary models for very low-mass stars of \citet{fernandes:2019}.
We used as inputs the stellar luminosity implied by $F_\mathrm{bol}$ and the \textit{Gaia} parallax, $L_\star = 5.92 \pm 0.07 \times 10^{-3}\,L_\sun$, and the metallicity inferred from the spectral analysis and SED fitting.
We assumed an age of $\gtrsim4$\,Gyr based on the optical spectral analysis.
This analysis gives $R_\star = 0.234 \pm 0.010\,R_\sun$, $T_\mathrm{eff} = 3300 \pm 50$\,K, and $\log g = 5.05 \pm 0.05$, all in excellent agreement with the SED-derived parameters.

Concerning the stellar mass, the evolutionary modeling gives $M_\star = 0.220 \pm 0.012\,M_\sun$, while the estimate from the empirical mass-$M_K$ relation of \citet{mann:2019} gives $M_\star = 0.243 \pm 0.012\,M_\sun$. We decided to merge these values with the procedure described in \cite{2018ApJ...853...30V} and obtained $M_\star = 0.231 \pm 0.012\,M_\sun$ as our best estimate for the stellar mass of TOI-2096.

In Table~\ref{table:stellar}, we summarize the stellar properties found in this subsection, which we use in the global model to derive the planetary parameters. In particular, taking these values of the stellar mass and radius, we find a stellar density of $\rho_{\star}=25.0\pm 4.1$~g/cm$^{3}$. We use this value as a prior for the global model, the posteriors of which yield a value of $\rho_{\star}=23.9\pm 3.3$~g/cm$^{3}$ (see Sect.~\ref{sec:transits} for details).

\begingroup
\begin{table}
\begin{center}
\renewcommand{\arraystretch}{1.15}
\begin{tabular*}{\linewidth}{@{\extracolsep{\fill}}l c c}
\toprule
Property & Value & Source \\
\midrule
Sp.\ type & M4 & Optical spectra \\[0.1cm]
$\mathrm{T_{eff}}$ (K) & $3300 \pm 50$ & SED \& EV$^{a}$ \\[0.1cm]
$\mathrm{[Fe/H]}$ & $-0.03 \pm 0.20$ & Opt.\ \& IR spectra \\[0.1cm]
$\mathrm{M_\star}$ ($\mathrm{M_\odot}$) & $0.231 \pm 0.012$ & Sect.~\ref{sec:stellar_cha} \\[0.1cm]
$\mathrm{R_\star}$ ($\mathrm{R_\odot}$) & $0.235 \pm 0.011$ & SED \\[0.1cm]
$\mathrm{F_{bol}}$ ($10^{-11}$ $\mathrm{erg\,s^{-1}\,cm^{-2}}$) & $8.066 \pm 0.093$ & SED \\[0.1cm]
$\mathrm{L_\star}$ ($10^{-3}$ $\mathrm{L_\odot}$) & $5.92 \pm 0.07$ & SED \\[0.1cm]
$\mathrm{\log\,g}$ & $5.05 \pm 0.05 $ & $\mathrm{M_\star}$, $\mathrm{R_\star}$ \\[0.1cm]
$\mathrm{\rho_\star}$ ($\mathrm{g\,cm^{-3}}$) & $23.9\pm 3.3$ & fit$^{b}$ \\[0.1cm]
Age (Gyr) & $\gtrsim$4 & Optical spectra \\[0.1cm]
Distance (pc) & $48.46\pm0.04$ & Parallax$^{c}$ \\[0.1cm]
\hline
\end{tabular*}
\end{center}
\caption{Derived properties of the host star. $^a$ from evolutionary models. $^b$ Fitted using a prior derived from the radius and mass (see Sect.~\ref{sec:transits}). $^c$ Computed from the parallax provided by \cite{gaia:2021_EDR3}.} 
\label{table:stellar}
\end{table}
\endgroup

\begingroup
\begin{table}
\begin{center}
\renewcommand{\arraystretch}{1.15}
\begin{tabular*}{\linewidth}{@{\extracolsep{\fill}}l c c}
\toprule
Parameter & Value & Source \\
\midrule
\multicolumn{3}{c}{\textit{Target designations}} \\
TIC      & 142748283           & 1 \\
2MASS    & J10062860+7449391   & 2 \\
{\it Gaia\/} EDR3 & 1126422935076082176 & 3 \\
WISE     & J100628.52+744938.3 & 4 \\
UCAC 4   & 825-014225          & 5 \\
\midrule
\multicolumn{3}{c}{\textit{Photometry}} \\
$TESS$	& 13.476 $\pm$ 0.007   & 1 \\
$B$	    & 17.78 $\pm$ 0.42     & 6 \\
$V$	    & 15.81 $\pm$ 0.32     & 6 \\
$Gaia$	& 14.7736 $\pm$ 0.0004 & 3 \\
$J$	    & 11.877 $\pm$ 0.021      & 2 \\
$H$	    & 11.298 $\pm$ 0.019      & 2 \\
$K$	    & 11.021 $\pm$ 0.017      & 2 \\
WISE 3.4 $\mu$m	& 10.873 $\pm$ 0.023 & 4 \\
WISE 4.6 $\mu$m	& 10.673 $\pm$ 0.021 & 4 \\
WISE 12 $\mu$m	& 10.495 $\pm$ 0.077 & 4 \\
WISE 22 $\mu$m	& 9.390 $\pm$ -- & 4 \\
\midrule
\multicolumn{3}{c}{\textit{Astrometry}} \\
RA  (J2000) & 10 06 28.58  & 3 \\
DEC (J2000) & +74 49 39.12  & 3 \\
RA PM (mas/yr)  & -25.826 $\pm$ 0.020  & 3 \\
DEC PM (mas/yr) & -75.104 $\pm$ 0.026   & 3 \\
Parallax (mas) & 20.636 $\pm$ 0.019 & 3 \\
\bottomrule
\end{tabular*}
\end{center}
\caption{
TOI-2096 stellar astrometric and photometric properties. 1. \citet{stassun2018}, 2. \citet{cutri:2003}, 3. \citet{gaia:2021_EDR3}, 4. \citet{cutri:2014}, 5. \citet{zacharias:2012}, 6. \citet{lasker:2008}.
\label{tab:starlit}} 
\end{table}
\endgroup

\section{Target vetting tests}\label{sec:validation}
\subsection{TESS pipeline data validation}
As described in Sect.~\ref{subsec:tess}, the SPOC pipeline extracted the photometry for TIC~142748283 using the 2-min data from the
three relevant sectors of the primary mission (14, 20, and 21), yielding two periodic transit-like signatures with periods of 3.119~d and 6.387~d. The TOI vetting team at MIT reviewed the SPOC Data Validation reports \citep{twicken2018,2019PASP..131b4506L} for this target and released it as TOI-2096 on 15 July 2020 \citep{guerrero2021}. The TOI report states that TOI-2096.01 and TOI-2096.02 have periodic transit signals with S/N of 7.0 and 10.4, respectively. While TOI-2096.02 passed all validation tests, TOI-2096.01, due to its shallower transits, narrowly failed some checks, falling just below the 7.1 threshold for the model fitter and the bootstrap test. This transit also failed the ghost diagnostic test \citep{twicken2018}. The star has two faint neighbors within 1\,arcmin, but both candidates passed the centroid tests. 
Therefore, we speculated that TOI-2096 is an actual multiplanet system and decided to trigger a follow-up campaign from the ground as described in Sect.~\ref{sec:gbobservations}.

\subsection{Seeing-limited photometry -- SG1}
\label{sec:sg1}

From \textit{Kepler} studies, it was found that multicandidate systems have a higher probability of being real planets \citep[see, e.g.,][]{lissauer:2012}, which favors our planetary interpretation for the two signals detected in TOI-2096 by the SPOC pipeline. However, TESS' pixel scale (21$\arcsec$) is much larger than Kepler's (4$\arcsec$), and TESS' PSF might be as large as 1$\arcmin$. These two factors increase the probability of contamination by a nearby eclipsing binary (NEB) \cite[see, e.g.,][]{kostov2019}. Indeed, deep periodic eclipses in a faint NEB might mimic the shallow transits observed on the target star due to dilution. Hence, exploring potential contamination from neighbor stars to confirm transit events on the star of interest is a must.  

In this context, we searched for potential false positives due to NEBs up to 2.5$\arcmin$ away from TOI-2096. This effort was made 
through the TFOP SG1, where the target star was observed with ground-based facilities at predicted transit times. The list of facilities involved in this task is summarized in Table~\ref{tab:GBobservations}. 

We confirmed the two signals on the target star at predicted times and ruled out NEBs, which could mimic the transits. In addition, observations in different filters showed nonchromatic dependence, which strengthens the hypothesis that the candidates are real planets.

\subsection{Archival imagery}

\begin{figure*}
    \centering
    \includegraphics[width=\textwidth]{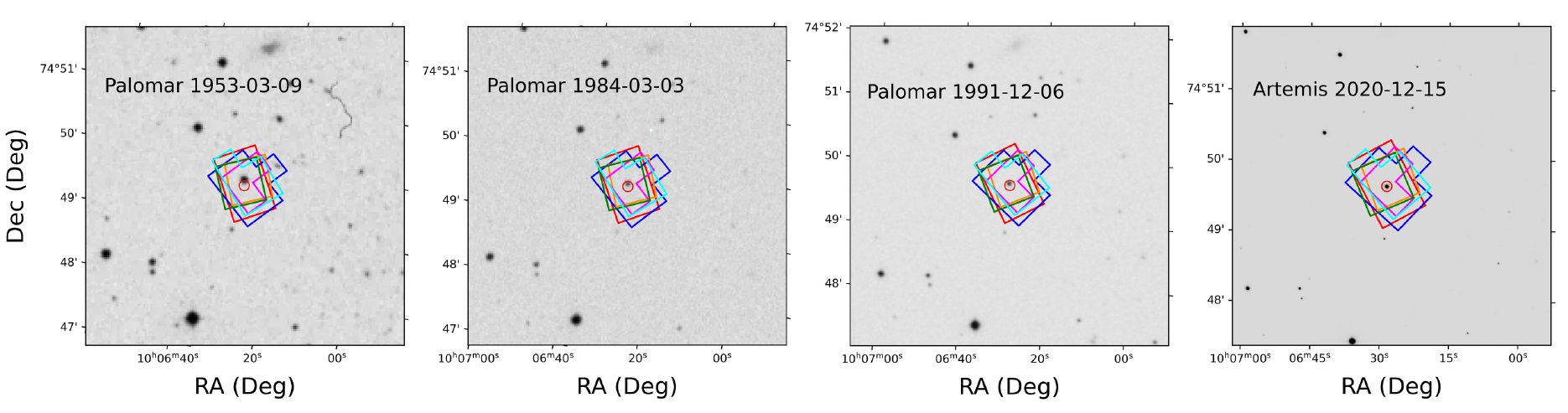}
    \caption{Archival images around TOI-2096 with TESS's apertures used in all the available sectors superimposed to assess for current, unresolved blending. From left to right: (1) 1953-03-09, (2) 1984-03-03, (3) 1991-12-06, and (4) 2020-12-15 archival. The red circle marks TOI-2096's location at the 2020 epoch.}
    \label{fig:poss}
\end{figure*}

We used archival images to rule out the presence of eclipsing binaries blended with TOI-2096 at its present-day location, which otherwise might be introducing the transit-like signals detected in our data \citep[see, e.g.,][]{quinn2019}. TOI-2096 has a moderately low proper motion of $\sim$0.08$\arcsec$/yr (see Table~\ref{table:stellar}). This projected angular velocity translated into a displacement of 5.3$\arcsec$ since the oldest archival image we found, taken from Palomar observatory in 1953 (see Fig.~\ref{fig:poss}), which has a pixel scale and PSF of 1.7$\arcsec$/pixel and 6.8$\arcsec$, respectively. Hence, while it seems that there is no star in the background, we can not confidently rule out the presence of a contaminating star. To this end, we should wait for a separation larger than the 1953 PSF; for example, 10$\arcsec$ would be enough \cite[see, e.g.,][]{schanche2022}. In such a case, we would need to wait until 2078.
On the other hand, in Sect.~\ref{sec:hri}, we described the high-resolution image that we obtained in February 2021 using the 'Alopeke speckle on Gemini North. This instrument provides a PSF $\sim$0.1$\arcsec$. Then, in a new observation in about three years, the star's displacement will be $\sim$0.24$\arcsec$, enough to disentangle if there is any blended background star.

\subsection{Statistical validation}

To validate the two planets in the system, we utilized the \triceratops package, developed specifically to aid in validating TESS candidates \citep{Giacalone2021}. In short, \triceratops is a Bayesian tool that uses knowledge of the transit, the target and the surrounding stars from \textit{Gaia} DR2, and the current understanding of planet occurrence rates to assess transit probabilities for several possible astrophysical false-positive scenarios. The tool can be used with TESS data alone or jointly with high-resolution contrast imaging to provide further constraints. According to \cite{Giacalone2021}, the two parameters needed to elucidate if a given candidate is a likely planet or a false positive are the false positive probability (FPP) and the nearby false positive probability (NFPP). To establish a candidate as a validated planet, the values of these two parameters should be FPP$<0.015$ and NFPP$<10^{-3}$. 

We used \triceratops to compute the FPP and NFPP values corresponding to the planet candidates in the system using the six sectors of TESS data. We found FPP values of $0.1133 \pm 0.0044$ and $0.1262 \pm 0.0034$ for planets b and c, respectively. It is worth noting that the NFPP captures the probability that the observed transit originates from a resolved nearby star rather than the target star; however, thanks to our ground-based follow-up, we know that the transits come from the target star. Hence, NFPP is 0. Incorporating the 'Alopeke contrast curves (see Sect.~\ref{sec:hri} and Fig.~\ref{fig:hri}) in the calculation reduces the FPP values to $0.0078 \pm 0.002$ and $0.0159 \pm 0.0013$. Moreover, as TOI-2096 is a multicandidate system, we can further constrain the FPP by applying the "multiplicity boost" described in  \cite{lissauer:2012}. Using the updated "boost" derived from TESS candidates for planets <6~R$_\oplus$ \citep{guerrero2021} brings the FPP estimate down to $(1.5\pm 0.37)\times 10^{-4}$ and $(3.0 \pm 0.25)\times10^{-4}$. These very low FPPs confidently rule out false-positive scenarios and provide a robust statistical validation for the system.

\section{Transit analysis}\label{sec:ta}
\subsection{Derivation of the system parameters}
\label{sec:transits}

We carried out our light-curve analyses using the \allesfitter package \citep{gunther2021}, which allows us to model planetary transits
using the \ellc package \citep{maxted2016} while accounting for other phenomena such as stellar flares, spots, and variability.
\allesfitter also allows for several ways to model the correlated noise, including polynomials, splines, and Gaussian Processes \citep[GPs;][]{rasmussen2004}, which are implemented through the \celerite package \citep{foreman2017,foreman2017b}. The parameters of interest are retrieved using a Bayesian approach implementing a Markov Chain Monte Carlo method \citep[see, e.g.,][]{hastings1970,ford2005} using the \emcee package \citep{foreman2013}, or the Nested Sampling inference algorithm \citep[see, e.g.,][]{feroz2009,feroz2019} using the \dynesty package \citep{speagle2020}. We used the Dynamic Nested Sampling algorithm to directly estimate the Bayesian evidence in this study. This strategy allows us to robustly compare a diverse set of orbital configurations.

For each planet, the fitted parameters were the ratio of planetary radius over stellar radius ($R_{p}/R_{\star}$), the sum of the stellar and planetary radius scaled to the orbital semi-major axis ($(R_{p}+R_{\star})/a$), the cosine of the orbital inclination ($\cos~i_{p}$), the mid-transit time ($T_{0}$), the orbital period ($P$), and, when considering eccentric scenarios, jointly the eccentricity and the argument of pericenter ($\sqrt{e}\cos \omega$ and $\sqrt{e}\sin \omega$).
Besides the physical parameters, we used GPs to model the correlated noise using the Matérn 3/2 kernel, which can describe smooth long-term trends and short-term stochastic variations by its two hyper-parameters: the amplitude scale ($\sigma$) and length scale ($\rho$). In addition, we added an error-scaling parameter for the white noise as a fitting parameter.

To reduce the number of free parameters in the models, we fixed the quadratic limb-darkening (LD) coefficients during the fitting process \citep[see, e.g.,][]{kipping2017,gunther2019}. To this end, we first computed the values of the LD coefficients in the physical $\mu$-space, $u_1$ and $u_{2}$, for each bandpass used in the data interpolating from the tables of \cite{claret2011}, using the effective stellar temperature ($T_\mathrm{eff}$), surface gravity ($\log g$), and metallicity ([Fe/H]). For the non-standard $I+z$ filter, we took the averages of the values for the standard filters $Ic$ and Sloan-z$'$. Then, following the relations of \cite{kipping2013}\footnote{$q_{1}=(u_{1}+u_{2})^{2}$ and $q_{2}=0.5u_{1}/(u_{1}+u_{2})$}, we converted $u_1$ and $u_{2}$ to the transformed $q$-space, $q_{1}$ and $q_{2}$, as required by \allesfitter. We report these values in Table~\ref{table:LD}. 

We followed the procedure described in \cite{gunther2019} to model our data. This procedure consists of several stages, where before performing a global model accounting for all the available data, we analyzed each light curve independently to estimate the GP hyper-parameters and the white noise parameter. The stages we followed were:

\begin{enumerate}

\item For the TESS data, we first recovered the TOIs through the \tls package \citep{tls}. Both planets TOI-2096\,b and c were successfully found, confirming the findings of the SPOC pipeline.
We took the orbital periods and transit times from \tls to refine the transit locations by conducting a preliminary fit of all the TESS sectors using wide, uniform priors.

\item We masked eight-hour windows around every transit midpoint found in 1) and fit for the noise and GP hyper-parameters in the out-of-transit data using wide uniform priors.

\item  We refined the planetary and orbital parameters, propagating the out-of-transit posteriors from 2) as priors into a fit of the full TESS data using Gaussian distributions. Planetary and orbital parameters were sampled from wide uniform priors.  

\item  For each ground-based observation, we estimated the white noise and the GP hyper-parameters from a wide uniform distribution by fixing the periods and the $R_{p}/R_{\star}$ values found in 3) and sampling the mid-transit times from a uniform distribution of one hour around the predicted mid-transit times.

\item  Unfortunately, some of the follow-up observations showed a high level of red noise. To estimate which are helpful to refine the transit parameters and which are dominated by red noise on scales larger than the transit signals, we fit each data set independently with two models: a pure-noise model and a transit-and-noise model. We recorded the Bayesian evidence in each case.
\begin{enumerate}
\item Pure-noise model: We fit the light curves by fixing the $R_{p}/R_{\star}$ to 0 (no planet is transiting in the data) and used the posteriors from 4) propagating them using Gaussian distributions. 

\item Transit-and-noise model: We fit the planets and orbital parameters by using the posteriors from 3) sampling them from uniform distributions and propagating the posteriors found in 4) using Gaussian distributions.
\end{enumerate}
\item For each follow-up observation, we computed the Bayes factor as $\Delta \ln Z = \ln Z_\mathrm{transit} - \ln Z_\mathrm{noise}$. There is strong evidence for a transit signal in the data when $\Delta \ln Z > 2.3$ \citep{kass1995}. Then, we only included the ground-based observations that fulfill such criteria for the global analyses. The full list of follow-up observations with their corresponding $\Delta \ln Z$ values is displayed in Table~\ref{tab:GBobservations}.                    

\item Finally, we conducted global model fits, including the TESS data and the ground-based observations that overcame stage 6). These global models correspond to different orbital configurations, for which we recorded the Bayesian evidence. In these global models, we used wide uniform priors for the planetary and orbital parameters and Gaussian distributions for the noise and GP baseline. In particular, the models corresponded to the following: 
\begin{enumerate}
\item Circular orbits for both planets; 
\item Eccentric orbit for TOI-2096\,b and circular orbit for TOI-2096\,c; 
\item Circular orbit for TOI-2096\,b and eccentric orbit for TOI-2096\,c; and
\item Eccentric orbits for both planets.
\end{enumerate}
\end{enumerate}
During the modeling procedure, we passed the stellar mass and radius obtained from our stellar characterization (see Sect.~\ref{sec:star}) as input into \allesfitter
to compute a normal prior on the stellar density of $\rho_{\star}=25.0\pm 4.1$~g/cm$^{3}$. During the fitting process, the stellar density is calculated at each Nested Sampling step from the fitted parameters via $\rho_{\star}\approx\frac{3\pi}{GP^{2}}(\frac{a}{R_{\star}})^{3}$ \citep{seager2003} and compared with the value provided in the prior. The fits are penalized when these two values disagree.        

\begin{figure*}
    \centering
    \includegraphics[width=0.999\textwidth]{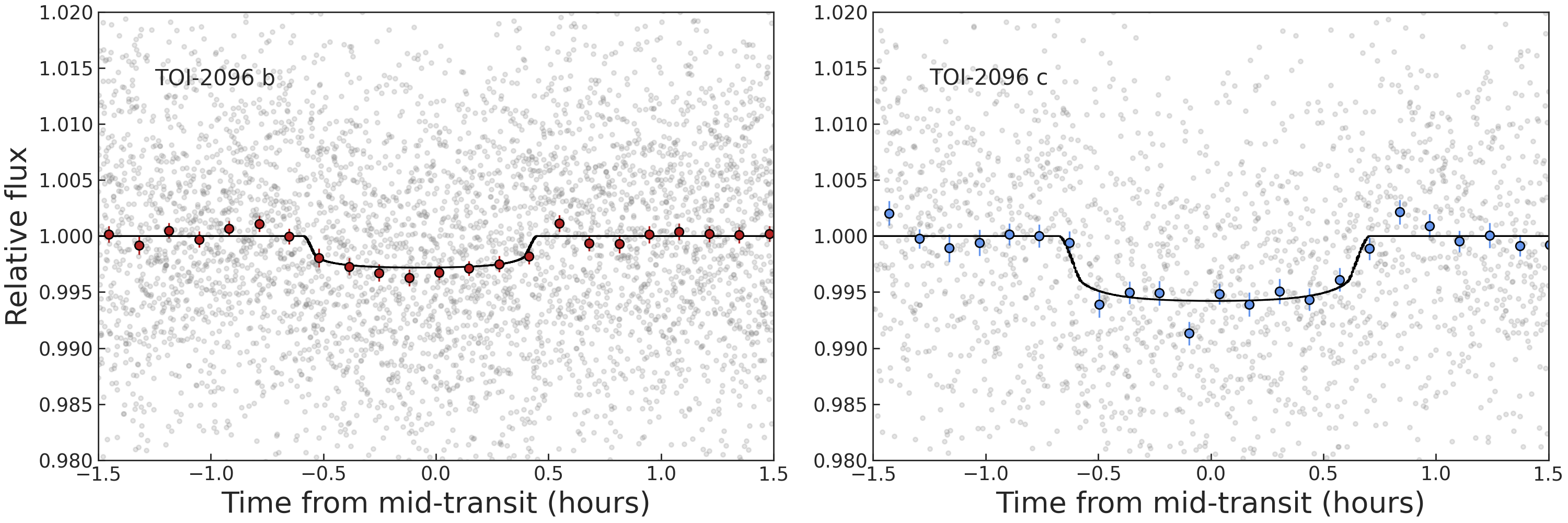}
    \caption{Detrended TESS photometry phase-folded to the 3.12~d period of TOI-2096\,b (left) and to the 6.39~d period of TOI-2096\,c (right) along with the best-fit transit model (solid black line). The unbinned data points are depicted in grey, while the colored circles with error bars correspond to 8-min bins.}
    \label{fig:lc_tess}
\end{figure*}

\begin{figure*}
    \centering
    \includegraphics[width=0.85\textwidth]{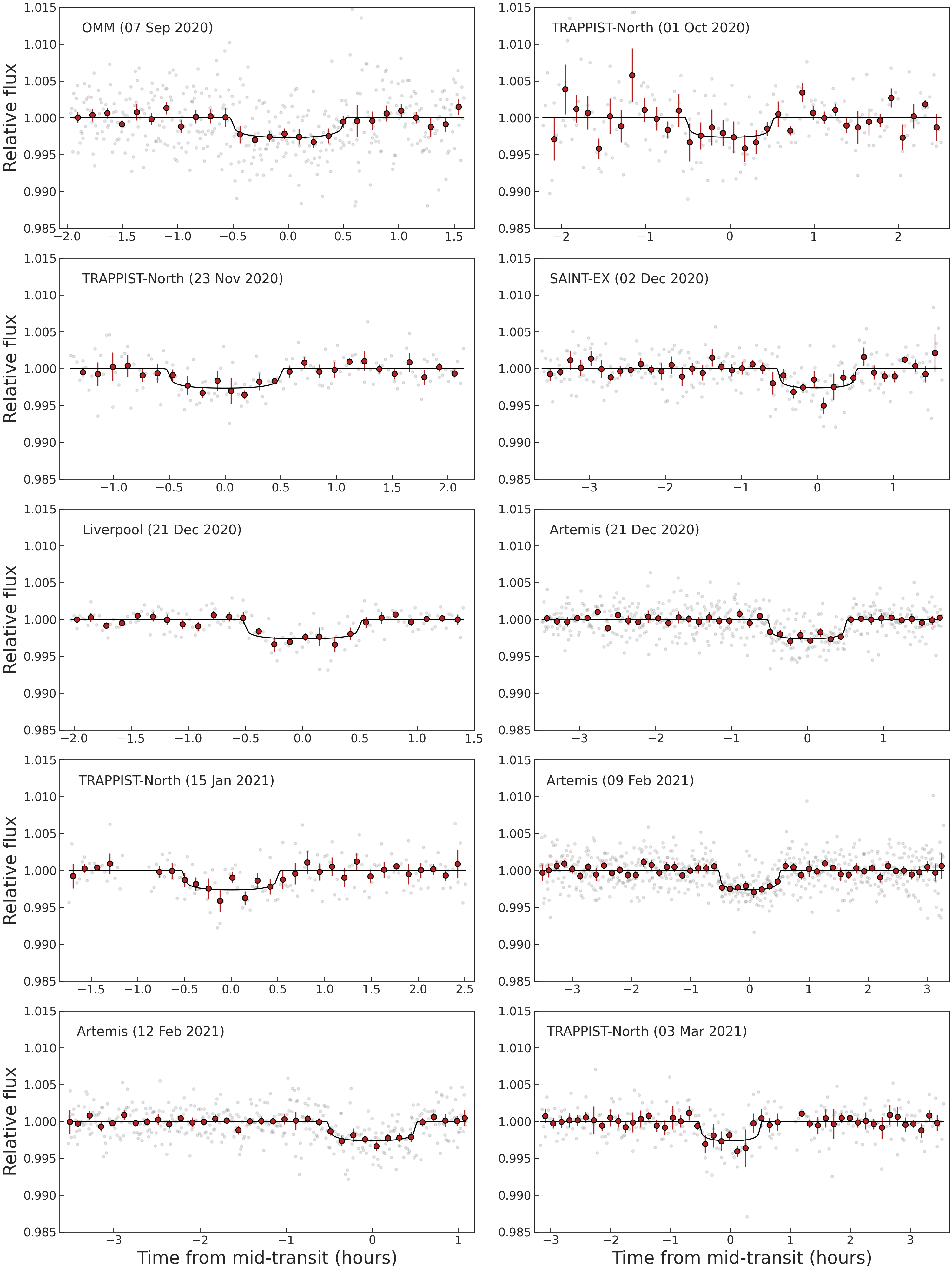}
    \caption{Phase-folded and detrended ground-based photometry of TOI-2096\,b transits along with the best-fit transit model (solid black line). The unbinned data points are shown in grey, while the red circles with error bars correspond to 8-min bins.}
    \label{fig:lc_toi2096b}
\end{figure*}

\begin{figure*}
    \centering
    \includegraphics[width=0.90\textwidth]{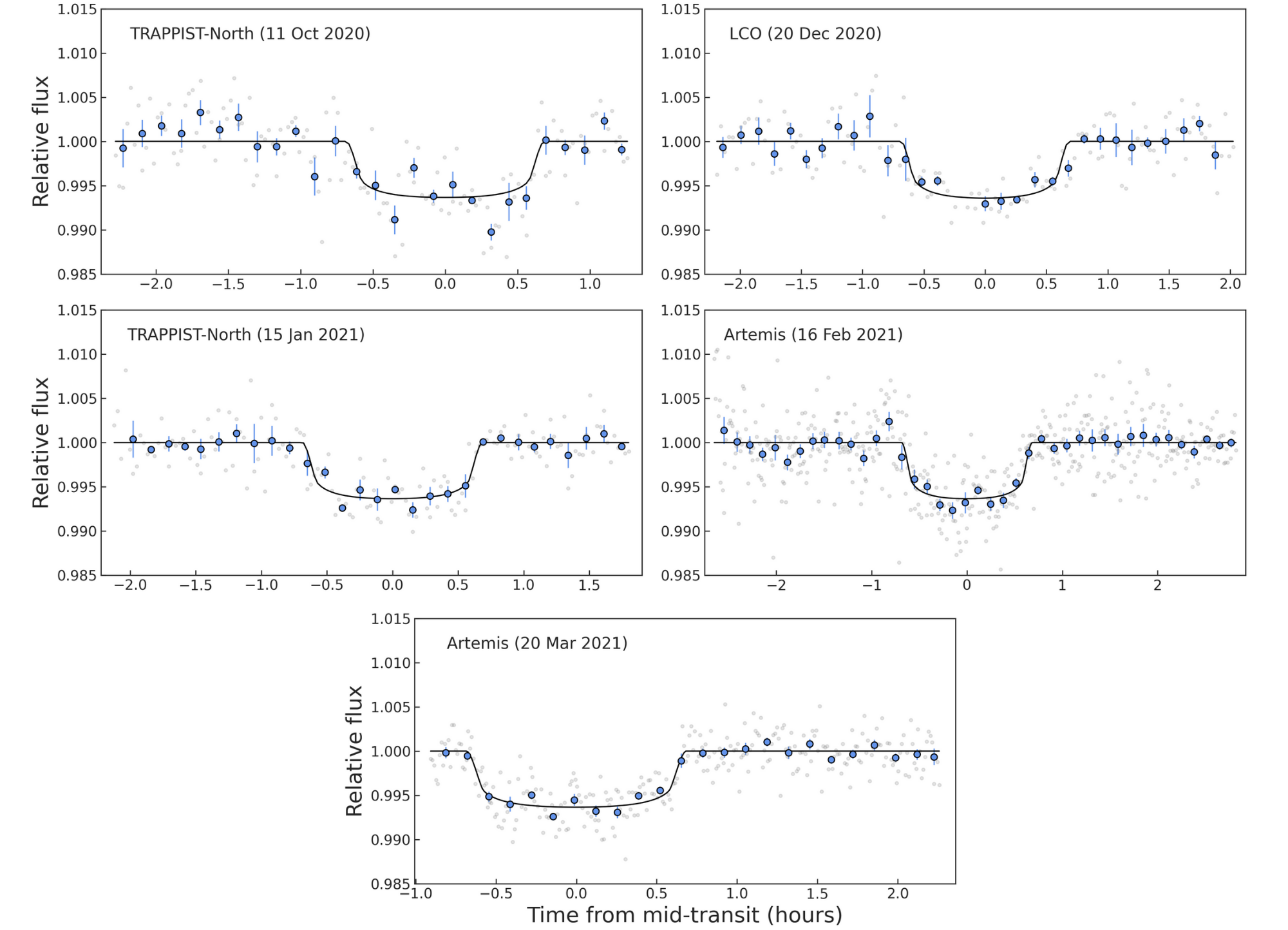}
    \caption{Phase-folded and detrended ground-based photometry of TOI-2096\,c transits along with the best-fit transit model (solid black line). The unbinned data points are shown in grey, while the blue circles with error bars correspond to 8-min bins.}
    \label{fig:lc_toi2096c}
\end{figure*}

Our modeling allowed us to determine that all the scenarios that include eccentric orbits were favored over the simplest one where both planets reside in circular orbits, model (a). On the other hand, models (b), (c), and (d) are all statistically indistinguishable (see Table~\ref{table:model_comparison}), and the four models considered here yielded nearly identical results within 1$\sigma$. These results hinted that likely the orbits of both planets are slightly eccentric; however, with the current data set, their values remain highly undetermined, requiring RV measurements to break this degeneracy and confirm this result. The resulting light curves from our model (d), with both planets having slightly eccentric orbits, are shown in Fig.~\ref{fig:lc_tess} (both planets in TESS data), Fig.~\ref{fig:lc_toi2096b} (planet TOI-2096\,b using ground-based photometry), and Fig.~\ref{fig:lc_toi2096c} (planet TOI-2096\,c using ground-based photometry). The fitted, derived, and predicted physical parameters are reported in Table~\ref{table:planet_params}, and the posterior distributions for the fitted physical parameters are displayed in Fig.~\ref{fig:posteriors}. The resulting posteriors gives $\rho_{\star}=23.9\pm 3.3$~g/cm$^{3}$, agreeing with the prior.

\begin{table*} 
\centering
\begin{threeparttable}
\begin{tabular}{l c c r}
\toprule
Parameter & Unit & TOI-2096\,b & TOI-2096\,c  \\
\midrule
\multicolumn{4}{c}{\it{Fitted Parameters}} \\
\midrule
$R_p / R_\star$ &  &  $0.0485^{+0.0018}_{-0.0017}$ & $0.0747^{+0.0012}_{-0.0011}$ \\ \smallskip
$(R_\star + R_p)/a_p$ & &  $0.0455^{+0.0023}_{-0.0020}$ & $0.0288^{+0.0014}_{-0.0012}$ \\ \smallskip
$\cos{i_p}$ & & $0.0132\pm0.0085$ & $0.0077^{+0.0041}_{-0.0045}$\\ \smallskip
Orbital period,  $P$ & days &  $3.1190633^{+0.000010}_{-0.0000093}$ & $6.387840\pm0.000012$ \\ \smallskip
Mid-transit time, $T_0$ & BJD$_{TDB}$-2457000 &  $2146.45853^{+0.00037}_{-0.00041}$ & $2147.45983\pm0.00034$ \\ \smallskip
$\sqrt{e}cos\omega$ & &  $0.05\pm0.43$ &  $0.00\pm0.38$ \\ \smallskip
$\sqrt{e}sin\omega$ & &  $-0.16^{+0.20}_{-0.22}$ & $-0.05\pm0.17$ \\ 
\midrule
\multicolumn{4}{c}{\it{Derived Parameters}} \\
\midrule
Planet Radius, $R_p$ & $R_\oplus$ & $1.243^{+0.077}_{-0.072}$ & $1.914\pm{0.095}$ \\ \smallskip
Orbital Eccentricity, $e$ & &  $0.15^{+0.27}_{-0.12}$ & $0.10^{+0.17}_{-0.06}$\\ \smallskip
Argument of Periastron, $\omega$ & $^{\circ}$ &  $245\pm{91}$ & $200\pm140$\\ \smallskip
Semimajor axis, $a$ & au &  $0.025 \pm 0.001$ & $0.040 \pm 0.002$\\ \smallskip
Inclination, $i$ & $^{\circ}$ &  $89.24\pm0.49$ & $89.56^{+0.26}_{-0.23}$\\ \smallskip
Transit duration, $T_{1-4}$ & hrs &  $1.072^{+0.024}_{-0.021}$ & $1.354^{+0.018}_{-0.016}$\\ \smallskip
$^{a}$ Equilibrium Temperature, $T_{eq}$ & K & $445 \pm 13$ & $349^{+10}_{-9}$\\ \smallskip
Transit Depth, $\delta$ & ppt &  $2.66\pm0.17$ & $6.36\pm0.18$\\ \smallskip
Impact parameter, $b$ & & $0.30^{+0.23}_{-0.19}$ &  $0.28^{+0.15}_{-0.16}$ \\ \smallskip
Insolation Flux, $S$ & S$_{\oplus}$ & $9.4\pm0.2$ & $3.7^{+0.1}_{-0.2}$  \\ 
\midrule
\multicolumn{4}{c}{\it{Predicted Parameters}} \\
\midrule
$^{b}$ Planet Mass, $M_p$ & $M_\oplus$ &  $1.9^{+1.4}_{-0.6}$ & $4.6^{+3.5}_{-1.8}$\\ \smallskip
$^{c}$ RV Semi-amplitude, $K$ & $m\, s^{-1}$ & $2.3$ & $3.7-9.0$  \\ \smallskip
$^{d}$ TSM &  & 6.0 & 54.0 \\ 
\midrule
\end{tabular}
\end{threeparttable}
\caption{Fit, derived, and predicted parameters for the TOI-2096 system. 
$^{a}$ Values derived using an albedo of 0.3 (Earth-like), and emissivity of 1. 
$^{b}$ Values estimated using the probabilistic mass--radius relationship implemented in the \forecaster package \citep{chen2017}. 
$^{c}$ Values obtained assuming rocky composition for planet b, and rocky and water-rich for planet c (see Sect.~\ref{sec:radialvelo}). 
$^{d}$ Transmission spectroscopy metric (TSM) from \citet{kempton_tsm} (see Sect.~\ref{sec:atmos}).}
\label{table:planet_params}
\end{table*}

\subsection{Check for transit chromaticity}
\label{sec:chromaticity}

As mentioned in Sect.~\ref{sec:sg1}, the early analyses of ground-based observations showed similar transit depths for different bands, that is, no evidence for a wavelength-dependent transit depth. To verify this behavior, we repeated the same global analysis as in Sect.~\ref{sec:transits}, corresponding to the model (d), but adding a free dilution term for each ground-based observation, while for the TESS data, the most informative data set, the dilution term was fixed to 0 \citep{gunther2021}. These dilution terms were sampled uniformly from -1 to 1. We verified that all the fitted dilutions for both planets are close to 0 at the 1$\sigma$ level. In addition, we compared the posterior distributions of the derived transit depths corresponding to the ground-based observations with the one from TESS, finding that all also agree at the 1$\sigma$ level (see Table~\ref{tab:depths}). Hence, we confirm that the transits for both planets do not show any chromatic dependence.

\begin{table}[hbt!]
\centering
\begin{tabular}{lcc}
\toprule
\toprule
\multirow{2}{*}{Bandpass} & Transit depth & Transit depth \\
 & TOI-2096\,b (ppt) & TOI-2096\,c (ppt) \\
\midrule
\vspace{0.1cm}
$i'$ & $2.71_{-0.16}^{+0.18}$ & $-$ \\
\vspace{0.1cm}
$I+z'$ & $2.68_{-0.18}^{+0.16}$ & $6.37_{-0.16}^{+0.17}$ \\
\vspace{0.1cm}
$z'$ & $2.65_{-0.17}^{+0.16}$ & $-$ \\
\vspace{0.1cm}
$SDSS-Z$ & $2.64_{-0.15}^{+0.18}$ & $-$ \\
\vspace{0.1cm}
$Ic$ & $-$ & $6.43_{-0.17}^{+0.19}$ \\
\vspace{0.1cm}
TESS & $2.69_{-0.17}^{+0.19}$ & $6.39_{-0.16}^{+0.17}$ \\
\bottomrule
\bottomrule
\end{tabular}
\caption{Transit depths returned by our global data analysis for which we allowed transit depth variations between the different bandpasses (see Sect. \ref{sec:chromaticity}).}
\label{tab:depths}
\end{table}

\subsection{Search for transit timing variations}
\label{sec:ttv1}

The resulting period ratio $P_{c}/P_{b}$ is 2.048, which places the system within $\sim$5$\%$ of the first-order 2:1 MMR. In such a situation, due to the gravitational interaction between the planets, one may expect some level of mutual orbital excitation, which in turn may induce measurable TTVs \citep{agol2005,holman2005}. The first-order MMRs are defined as the period ratio being close to $P_{in}/P_{out} \approx (i-1)/i$, whereby $i$ is an integer, and P$_{in}$ and P$_{out}$ are the periods of the inner and outer planets, respectively. Then, planets' mid-transit times show sinusoidal variations in a super-period timescale defined as $P_{TTV}=1/|i/P_{out} - (i-1)/P_{in}|$. In the case of the TOI-2096 system, we found a $P_{TTV}\sim$133~d.  

The amplitude of TTVs depends on the planetary masses and eccentricities, which are highly undetermined from our photometric analyses presented in Sect.~\ref{sec:transits}. Here, we aim to search for TTVs that allow us to constrain these two 
parameters. We repeated the global analysis carried out previously corresponding to the model (d), but this time we fixed the orbital periods and epochs to the values reported 
in Table~\ref{table:planet_params}, while allowing for each transit of the two planets a timing offset with respect to the linear prediction. 

We focussed our search for TTVs on ground-based photometry only. This strategy is motivated by the low S/N of individual TESS transits, which hinders precise measurement of their timings. Moreover, including TESS data would mean adding 47 (number of transits for planet b) and 23 (number of transits for planet c) extra free parameters to the 62 free parameters of the model (d) (see Table~\ref{table:model_comparison}), resulting in more than 130 free parameters. 
Neither MCMC nor Nested Sampling are suited to reliably account for all the covariances in such a high dimensionality \citep[see e.g.,][]{gunther2019}. We also removed from our TTVs search two ground-based observations: one corresponding to Artemis on December 12, 2020, because it offered a lower Bayes factor than the concurrent Liverpool observation (see Table~\ref{tab:GBobservations}), and the other corresponding to TRAPPIST-North on January 15, 2021, because a meridian flip happened some minutes before the ingress, affecting correct estimation of the transit mid-point measurement. 

\begin{figure}
    \centering
    \includegraphics[width=\columnwidth]{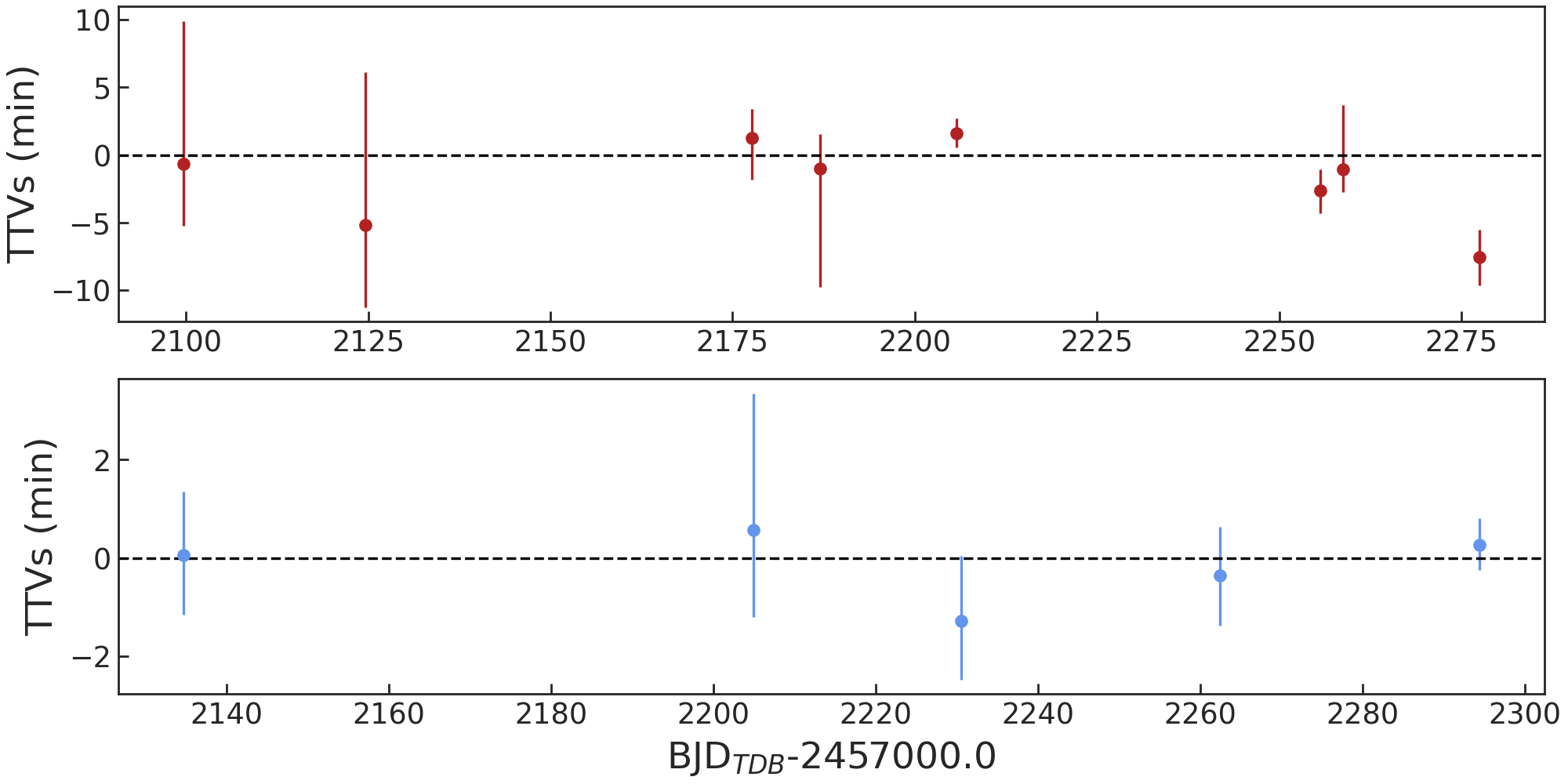}
    \caption{Transit timing variations as a function of time for TOI-2096\,b (top panel) and TOI-2096\,c (bottom panel). We used only the ground-based observations highlighted with $\ddagger$ in Table~\ref{tab:GBobservations} (see text for details).}
    \label{fig:ttvs}
\end{figure}

In total, we accounted for eight observations for TOI-2096\,b and five observations for TOI-2096\,c. The time baseline coverage provided by these ground-based observations lasted for 177~d for planet b and 160~d for planet c, which exceeds the $P_{TTV}$ of 133~d. The detection of the TTVs thus only depends on their amplitudes. In Table~\ref{tab:ttvs} are presented the timings and the corresponding TTVs that we found in our analysis, and they are displayed as a function of time in Fig.~\ref{fig:ttvs}. From these results, we conclude that while TTVs detection is challenging in the current data set, some points show marginally significant offsets from their linear predictions. These deviations hint that the TOI-2096 system might be suitable for an intensive high-precision photometric follow-up to accurately determine planetary masses and eccentricities. In Sect.~\ref{sec:dyn_ttv} we present some prospects for such future observations and follow-up.

\section{Planet searches and detection limits from the TESS photometry}\label{sec:search}

As stated in Sect.~\ref{sec:obs}, TESS observed TIC~142748283 in sectors 14, 20, 21, 40, 41, and 47 and issued two planetary candidates: TOI-2096\,b and TOI-2096\,c. However, 
more planets might remain unnoticed in the data due to the threshold of the SPOC and the Quick-Look Pipeline \citep[QLP; ][]{qlp2020a,qlp2020b,qlp2021} to trigger a detection alert (MES=7.1$\sigma$).

In this context, we searched for additional planetary candidates by means of our custom pipeline \sherlock\footnote{\sherlock (\textbf{S}earching for \textbf{H}ints of \textbf{E}xoplanets f\textbf{R}om \textbf{L}ightcurves \textbf{O}f spa\textbf{C}e-based see\textbf{K}ers) code is fully available on GitHub: \url{https://github.com/franpoz/SHERLOCK}}, originally presented by \cite{pozuelos2020} and used successfully in a number of studies \citep[see, e.g.,][]{demory2020,wells2021,schanche2022,sebastian2021,vangrootel2021}.

\sherlock is a dedicated pipeline that allows the exploration of space-based data to recover known planets and alerts, and to search for new signals attributable to planets.
The pipeline combines six different modules to (1) download and prepare the light curves from their online repositories, (2) search for planetary candidates, (3) perform a semi-automatic vetting of the interesting signals, (4) compute a statistical validation, (5) model the signals to refine their ephemerides, and (6) compute observational windows from ground-based observatories to trigger a follow-up campaign.

During step (1) \sherlock computes the Lomb--Scargle periodogram \citep{lomb1976,scargle1982,lsp2018} to identify any stellar variability, such as stellar rotation, that might hinder the identification of planetary transits. 
\sherlock extracts the value of the highest-power peak using the \lightkurve package \citep{lightkurve}, and the associated variability is corrected 
by fitting a sum of sines and cosines employing the \texttt{cosine} function provided by the \wotan package \citep{wotan2019}. In the case of TOI-2096, we found a soft-flux variation 
that hints at a stellar rotational period of 4.40~d. To optimize the transit search and remove any other undesired trends such as instrumental drifts, \sherlock uses a multidetrend approach, implemented via the \wotan package, whereby the nominal PDCSAP light curve is detrended several times using a biweight filter, by varying the window size. In our case, we performed 20 detrends with window sizes ranging from 0.20 to 1.30~d. Each new detrended light curve, jointly with the nominal PDCSAP flux, is processed by the \tls package \citep{tls}, which is optimized for detecting shallow periodic transits using an analytical transit model based on stellar parameters. This multidetrend approach is motivated by the associated risk of removing transit signals, particularly short and shallow signals. Hence, with this strategy, we converged on the most efficient detrend, which allowed us to recover the alerted planets and search for any potential extra ones with the best S/N and signal-detection-efficiency (SDE). 

\sherlock searches for signals by performing a search-find-mask loop; that is, once a signal is found, it is stored and masked, and then the search continues until no more signals are found above user-defined S/N and SDE thresholds. This looping process is performed up to five times because results found beyond five runs are less reliable due to the accumulated gaps in a given light curve after many mask-and-run iterations. We followed the strategy presented in \cite{delrez2022}, performing two independent transit searches for extra planets by considering all sectors simultaneously. First, we focused our search on orbital periods ranging from 0.5 to 40~d, where a minimum of two transits was required to claim a detection. We recovered the TOI-2096.01 and .02 signals in the first and second runs, respectively. In the subsequent runs, we did not find any other signal that hinted at the existence of extra transiting planets.
Second, we focused on longer orbital periods, ranging from 40 to 80 days, where single events could be recovered. This case also yielded negative results, with all the signals found being attributable to variability, noise, or systematics. 

The lack of extra signals in TESS data might be due to one of the following scenarios \cite[see, e.g.,][]{wells2021,schanche2022}: 
(1) no other planets exist in the system; (2) they do exist, but they do not transit; (3) they do exist and transit, but have orbital periods longer than the ones explored in this study; or (4) they do exist and transit, but the photometric precision of the data is not accurate enough to detect them. 

Scenarios (1) and (2) might be further explored by employing RV follow-up as discussed in Sect.~\ref{sec:radialvelo}. Scenario (3) can be tested by adding a longer time baseline, for example, using data from Sector 53 and the upcoming Sector 60. To evaluate scenario (4), we studied the detection limits of the TESS photometry performing injection-and-recovery experiments with the \matrixtk code  \footnote{{The \matrixtk (\textbf{M}ulti-ph\textbf{A}se \textbf{T}ransits \textbf{R}ecovery from \textbf{I}njected e\textbf{X}oplanets) code is open access on GitHub: \url{https://github.com/PlanetHunters/tkmatrix}}} \citep{matrix2022}. 

\matrixtk injects synthetic planets over the PDCSAP light curve that contains the six sectors used in this study. We explored the $R_{\mathrm{planet}}$--$P_{\mathrm{planet}}$ parameter space in the ranges of 0.5--3.0\,R$_{\oplus}$ with steps of 0.25\,R$_{\oplus}$, and 0.5--50.0 days with steps of 1.03 days. Moreover, for each combination of $R_{\mathrm{planet}}$--$P_{\mathrm{planet}}$ \matrixtk explores four different phases, that is, different values of $T_{0}$. In total, we explored 2,156 scenarios. For simplicity, the injected planets have impact parameters and eccentricities equal to zero. Once the synthetic planets are injected, \matrixtk detrends the light curves using a bi-weight filter with a window size of 1 day, which was found to be the optimal value during the \sherlock search and masked the transits corresponding to TOI-2096 b, and c. A synthetic planet is recovered when its epoch matches the injected epoch with 1~hour accuracy, and its period is within 5\,\% of the injected period. 
Since we injected the synthetic signals in the PDCSAP light curve, these signals were not affected by the PDCSAP systematic corrections; hence, the detection limits that we find are the most optimistic scenario \citep[see, e.g.,][]{pozuelos2020,eisner2020}.

\begin{figure}
\includegraphics[width=\columnwidth]{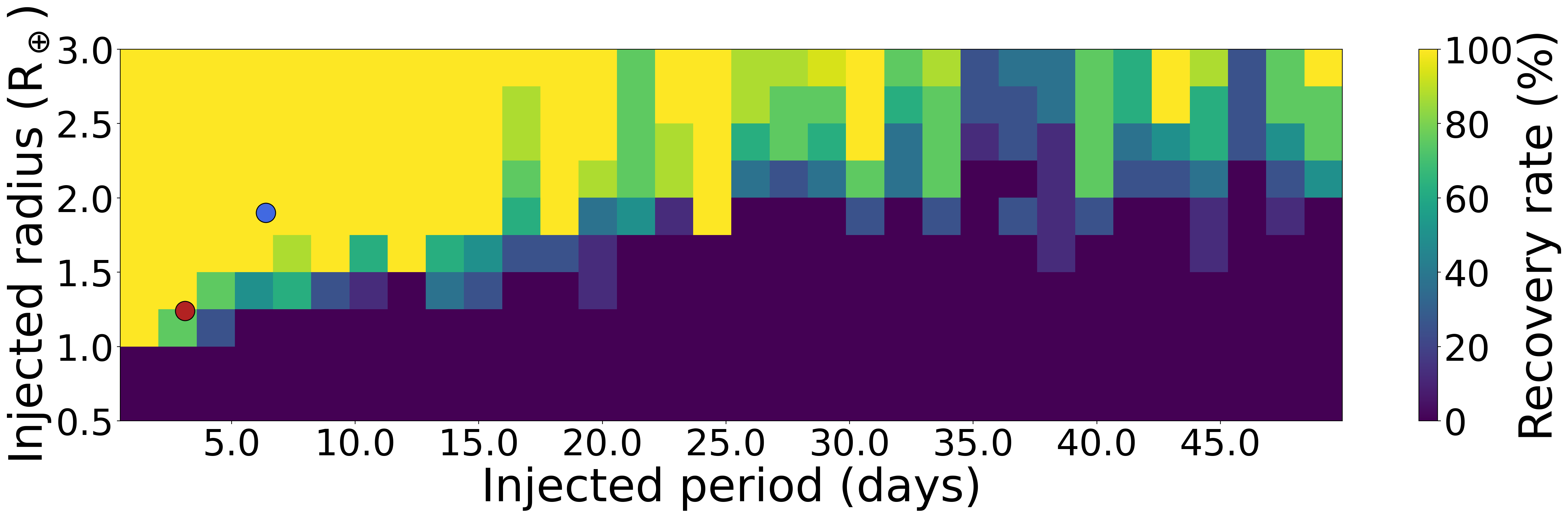}
\caption{Injection-and-recovery experiment performed to test the detectability of extra planets in the system using the six TESS sectors described in Sect.~\ref{sec:obs}.  We explored a total of 2156 different scenarios. Each pixel evaluated about 6 scenarios, that is, 6 light curves with injected planets having different $P_{\mathrm{planet}}$, $R_{\mathrm{planet}}$, and T$_{0}$. Larger recovery rates are presented in yellow and green colors, while lower recovery rates are shown in blue and darker hues. Planets smaller than 1.0~R$_{\oplus}$ would remain undetected for the explored periods. Red and blue dots refer to the planets TOI-2096\,b, and c, respectively.} \label{fig:recovery}
\end{figure}

The detectability map resulting from these injection-and-recovery experiments is shown in Fig.~\ref{fig:recovery}. We found that Earth- and sub-Earth size planets would remain unnoticed for the complete set of periods explored. For short orbital periods $<$20~days, the presence of planets larger than 2~R$_\oplus$ seems unlikely, with recovery rates from 80 to 100$\%$. For orbital periods $>$20~days, planets with sizes smaller than 2~R$_\oplus$ would remain undetectable (recovery rates $\sim 0\%$), while larger sizes would be challenging but possible to detect, with recovery rates >20$\%$. We note that planets TOI-2096\,b and c lie in regions with recovery rates of 80 and 100$\%$, respectively.

\section{Dynamical analysis}\label{sec:dyn}


\subsection{Dynamical stability}
\label{sec:stability}

From Sect.~\ref{sec:transits} we concluded that TOI-2096 is a system composed of two planets with sizes of $\sim$1.24~R$_{\oplus}$ (TOI-2096\,b) and $\sim$1.9~R$_{\oplus}$ (TOI-2096\,c), likely in eccentric orbits. For TOI-2096\,b, its size points to a terrestrial composition; however, for TOI-2096\,c, its size might correspond to either a terrestrial or volatile-rich planet \citep{otegi2020}.
Unfortunately, from our analyses, both eccentricity and mass, the two parameters with the highest impact on the orbital stability of planetary systems, are largely degenerate.

Here, we aim to find additional restrictions to these parameters based on dynamic stability. We evaluated the Mean Exponential Growth factor of Nearby Orbits parameter, $Y(t)$ \citep[MEGNO;][]{cincottasimo1999,cincottasimo2000,cincotta2003}, which is widely used to explore the stability of planetary systems     
\cite[see, e.g.,][]{hinse2015,Jenkins2019,delrez2021}. In particular, we used the MEGNO implementation provided by {\scshape rebound} \citep{rein2012}, an N-body integrator that employs the Wisdom-Holman WHfast code \citep{rein2015}. Its time-averaged mean value, $\langle Y(t) \rangle$, amplifies any stochastic behavior, which can be used to distinguish between quasi-periodic (if $\langle Y(t\rightarrow \infty)\rangle \rightarrow$ 2) or chaotic trajectories (if $\langle Y(t\rightarrow \infty)\rangle \rightarrow$ $\infty$).

In the first step, we explored if there is any stability limitation for the planetary masses and the orbital eccentricities, following \cite{delrez2021}---that is, by building stability maps
exploring the $M_{b}-M_{c}$ and $e_{b}-e_{c}$ parameter spaces. We varied the planetary masses from 1 to 10 M$_{\oplus}$ and eccentricities from 0.0 to 0.6 for both planets, taking 20 values of each range. Hence, each stability map has 20$\times20$ pixels. For each realization, we set ten different initial conditions randomly sampled by varying the argument of periastron and the mean anomaly from 0 to 2$\pi$. The final result stored for each pixel is the average of these ten different initial conditions. Therefore, each stability map consisted of 4,000 scenarios. For the stability map corresponding to the planet masses, we froze the planetary eccentricities to their nominal values reported by Table~\ref{table:planet_params}. On the other hand, when exploring the parameter space corresponding to the orbit eccentricities, we froze the planetary masses to their nominal values predicted by \cite{chen2017}. The integration time and time-step were 10$^{6}$ orbits of the outermost planet and 5$\%$ of the orbital period of the innermost planet, respectively.

We found that the system is fully stable over the whole range of masses explored with a mean $\Delta \langle Y(t) \rangle$= 2.0 $-$ $\langle Y(t) \rangle = 7\times10^{-4}$, so we cannot set extra constraints on planetary masses. On the other hand, from the 
eccentricity map, we found that the TOI-2096 system tolerates planetary eccentricities that satisfy the condition $e_{b}\leq0.27-1.50\times e_{c}$. This relationship implies that the maximum eccentricity for planet b (when planet c has a circular orbit) is 0.27, and the maximum eccentricity for planet c (when planet b has a circular orbit) is 0.18. Hence, the system might become unstable for eccentricity values that do not meet this condition. 

\begin{figure}
    \centering
    \includegraphics[width=\columnwidth]{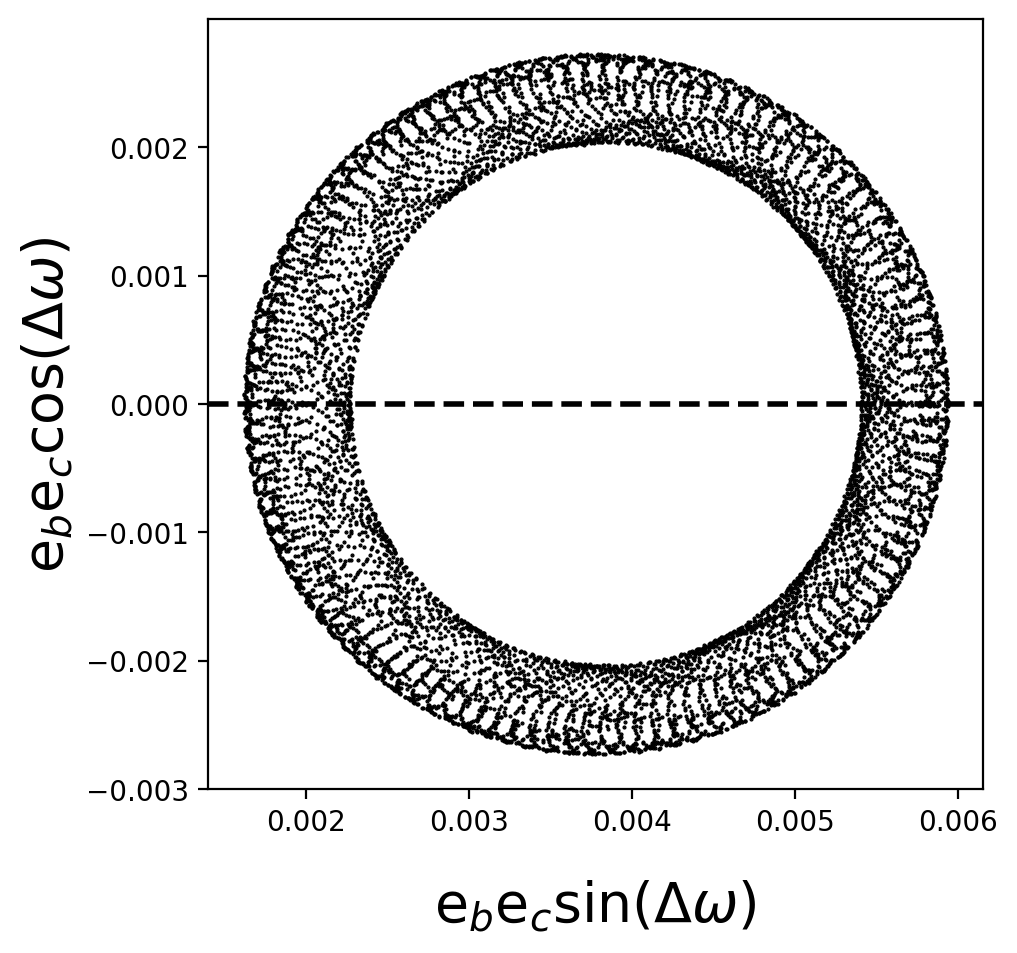}
    \caption{Polar plot of the apsidal trajectory ($e_b e_c$ versus $\Delta \omega$) for the b and c planets over a period of $10^6$ years. The figure shows that the apsidal modes are librating around the 2:1 MMR (see Sect.~\ref{sec:stability}).}
    \label{fig:epsilon}
\end{figure}

We further investigated the dynamical behavior of the planets to examine the effect of their lying close to the 2:1 MMR. To achieve this, we calculated the trajectory of the apsidal modes over a time period of $10^6$ years during the dynamical simulation, adopting the system parameters provided in Table~\ref{table:planet_params}. Apsidal motion in the context of interacting exoplanetary systems falls into categories of libration and circulation, and are separated by a boundary called a secular separatrix \citep{barnes2006a,barnes2006c,kane2014b,kane2019c}. The apsidal trajectories for planets b and c in the TOI-2096 system are represented graphically in polar form in Fig.~\ref{fig:epsilon}. The polar trajectories form an ellipse that does not encompass the origin, suggesting that the trajectories are librating. Furthermore, the trajectories lie entirely in the positive $e_b e_c \cos \Delta \omega$ direction, which is indicative of an aligned system configuration. The system is, therefore, in an aligned, stable libration around the 2:1 MMR.

\subsection{Transit timing variations}
\label{sec:dyn_ttv}

As stated previously in Sect.~\ref{sec:ttv1}, planets TOI-2096\,b and c are close to the 2:1 MMR, a configuration that might lead the planets to have measurable TTVs \citep[see, e.g.,][]{lithwick2012}. Indeed, our preliminary results from our search for TTVs hinted that some deviation from the linear ephemerides seems to exist. Unfortunately, the limited precision in our data set prevents us from constraining the planetary masses from transit times. 
Therefore, this subsection aims to provide a detailed analysis of the expected TTVs amplitudes for TOI-2096\,b and c, to elucidate if future follow-up campaigns might pin down their planetary masses. We followed a slightly different procedure to the one presented in \cite{cloutier2020}. In particular, our strategy consisted of generating 1,000 synthetic system configurations by drawing orbital periods and mid-transit times from the values reported in Table~\ref{table:planet_params} following normal distributions. We imported the mass distributions from the \texttt{Forecaster} package \citep{chen2017} for the planetary masses.
For the eccentricities, from our analyses carried out in Sect.~\ref{sec:transits}, while eccentric orbits are favored against circular, their values are poorly constrained. Recent studies suggest a relationship between the number of planets in systems and their eccentricities: the larger the number of planets, the lower the eccentricities \citep[see, e.g., ][]{limbach2015,zinzi2017,zhu2018}. For the particular case of two-planet systems, the distribution of eccentricities follows a lognormal distribution with $\mu$=-1.98 and $\sigma$=0.67 \citep{mathias2020}. We drew the eccentricities from such a distribution, adding the additional restriction found previously in Sect.~\ref{sec:stability} that limits the mutual values by $e_{b}\leq0.27-1.50\times e_{c}$.
Finally, we sampled the argument of periastron and the mean anomaly following a random distribution from 0 to 2$\pi$. We computed the TTVs amplitude value for each synthetic system using the \texttt{TTVFast2Furious} package \citep{hadden2019}. Considering 1,000 scenarios, we obtain the TTVs distributions for each planet. 
In this process, there is a non-null probability of drawing scenarios with very extreme values of masses and eccentricities at the same time, which was not considered before in Sect.~\ref{sec:stability}. These extreme cases may lead the system into unstable configurations. Then, as a sanity check, we evaluated the stability of each scenario by computing the MEGNO parameter for an integration time of 10$^{5}$ orbits of the outermost planet. To build the TTVs distributions, we only considered scenarios where $\Delta \langle Y(t) \rangle$= 2.0 $-$ $\langle Y(t) \rangle < 0.1 $. This restriction allows us to ensure that the drawn systems are realistic.  
In fact, we found that the drawn scenarios are highly stable, with 97.3$\%$ of them clustering around 1.9--2.1, supporting our results presented in Sect.~\ref{sec:transits} and the interpretation of TOI-2096 as a genuine planetary system \citep[see, e.g.,][]{chambers1996,barnes2004,fabrycky2014}.

\begin{figure}
    \centering
    \includegraphics[width=\columnwidth]{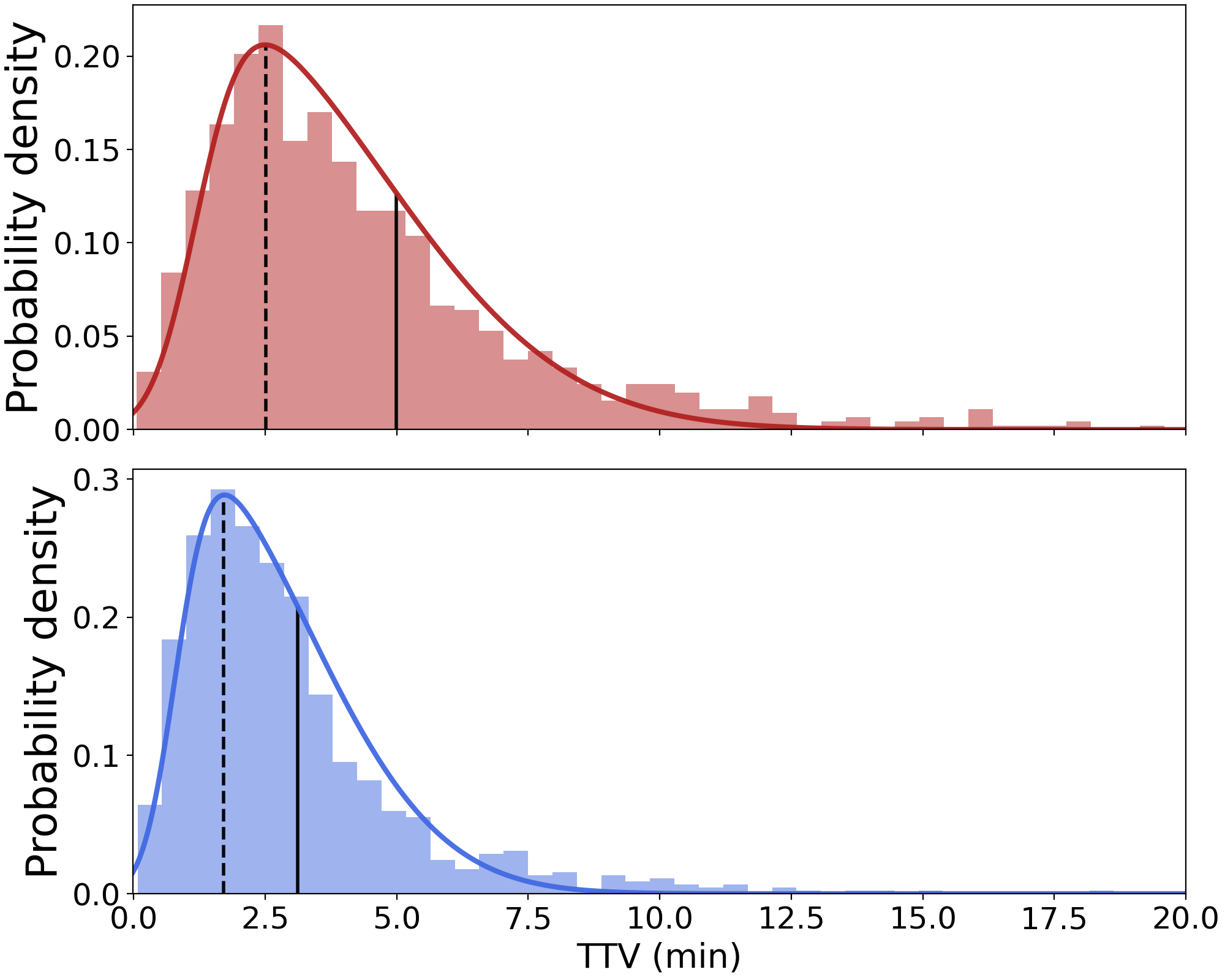}
    \caption{Expected TTVs amplitudes for planets TOI-2096\,b (upper panel) and TOI-2096\,c (lower panel). Dashed and solid vertical lines correspond to the mode and the mean of the PDF.}
    \label{fig:ttv_expected}
\end{figure}

We obtained that the TTVs for each planet follow a nonsymmetric normal distribution, which we fitted using the Skew-normal function from the \texttt{scipy} package \citep{scipy}. This package allowed us to derive each planet's TTVs probability density function (PDF). The results are displayed in Fig.~\ref{fig:ttv_expected}. For planets TOI-2096\,b and c, we found the mode of the PDF to be $\sim$2.5 and $\sim$1.7~min, respectively. On the other hand, we found the mean of the PDFs at $\sim$5~min and $\sim$3~min. Finally, with very low probabilities, the maximum TTVs found in our results were about $\sim$17~min and $\sim$13~min, respectively. Hence, while challenging, these results show that the TOI-2096 system might be suitable for measuring the planetary masses by performing an intense follow-up campaign using high-precision photometry, able to achieve a precision of $\lesssim$2 min in the transit mid-times. In our data set, on average, we got a precision of $\sim$5~min, with only a few observations reaching a precision of $\lesssim$2 min. In particular, for planet b, the Liverpool telescope yielded the best mid-transit time measurement, with a precision of 1 min, on December 21, 2021. On the other hand, for planet c, the best measurement was obtained using the Artemis telescope, providing a precision of 0.5 min, on March 20, 20221 (see Table~\ref{tab:ttvs}).  

\subsection{Stability of TOI-2096 with an additional planet}

In recent years, a high multiplicity of planets orbiting M-dwarfs stars has been found. For example, \cite{dressing2015} using a four-year Kepler dataset, reported a cumulative occurrence rate of 2.5$\pm$0.2~planets with sizes between 1--4~R$_{\oplus}$ per M dwarf. This occurrence rate is consistent with values obtained by combining radial velocities and microlensing detections, which yielded an estimation of 1.9$\pm$0.5 \cite[see, e.g.,][]{clanton2014}. Moreover, in a later revision of Kepler data, \cite{ballard2016} found that half of the M dwarfs are orbited by five or more planets on coplanar orbits. A review by \cite{tuomi2019} combining a large data set obtained by different instruments concluded that each M dwarf harbors at least 2.39$^{+4.58}_{-1.36}$~planets, with Earths, super-Earths, and mini-Neptunes residing in short orbital periods ranging from a few days to a hundred days. In this context, we explore under which conditions this two-planet system might dynamically accommodate a hypothetical third planet without perturbing the long-term stability. 

From Sect.~\ref{sec:search} we concluded that extra planets might exist and transit but remain unnoticed. For example, we found that an Earth-size planet would be undetectable within the current data set. 
We used the Stability Orbital Configuration Klassifier (SPOCK; \cite{spock2020}), a machine-learning model capable of classifying the stability of compact 3+ planetary systems over $10^{9}$ orbits of the innermost planet, which in this case translates into $\sim$10$^{7}$~yr. For planets TOI-2096\,b and c, we assumed the nominal values presented in Table~\ref{table:planet_params}. We injected the hypothetical third planet with its mass ranging from 0.5 to 4.0~M$_{\oplus}$ and semi-major axis from 0.001 to 0.06~au. For simplicity, we assumed the hypothetical third planet orbits in a circular orbit. We built a 300$\times$300 pixels stability map corresponding to 90,000 different scenarios. For each scenario, SPOCK computed the stability probability from 0 to 1, where 0 means a nonstable case and 1 is a fully stable configuration. 

\begin{figure}
    \centering
    \includegraphics[width=0.48\textwidth]{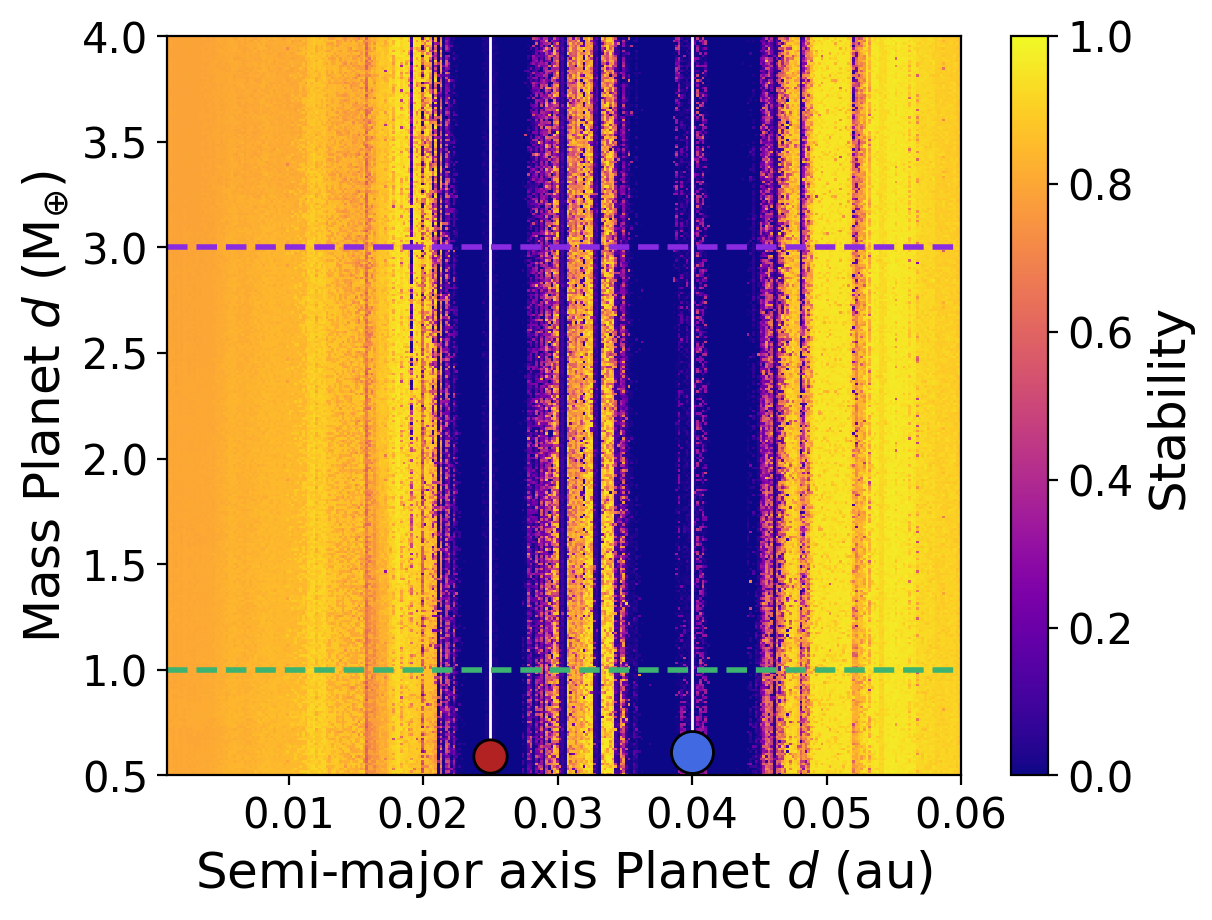}
    \caption{Stability map of TOI-2096 when a hypothetical third planet is injected into the system. The map contains 300$\times$300 pixels, in the M$_{d}$-- $a_{d}$ parameter space. The stability of each scenario is evaluated using the SPOCK package, establishing the level of stability of a given configuration with a value ranging from 0 (unstable) to 1 (fully stable). Then, the yellow regions refer to stable cases, while dark-blue hues are unstable. The horizontal dashed lines remark the particular cases of masses corresponding to 1.0 (green) and 3.0~M$_{\oplus}$ (purple), which are displayed in Fig.\ref{fig:3planet_extension}. The vertical white lines refer to the orbital locations of planets b (bottom red dot) and c (bottom blue dot).}
    \label{fig:3planet}
\end{figure}

\begin{figure}
    \centering
    \includegraphics[width=0.5\textwidth]{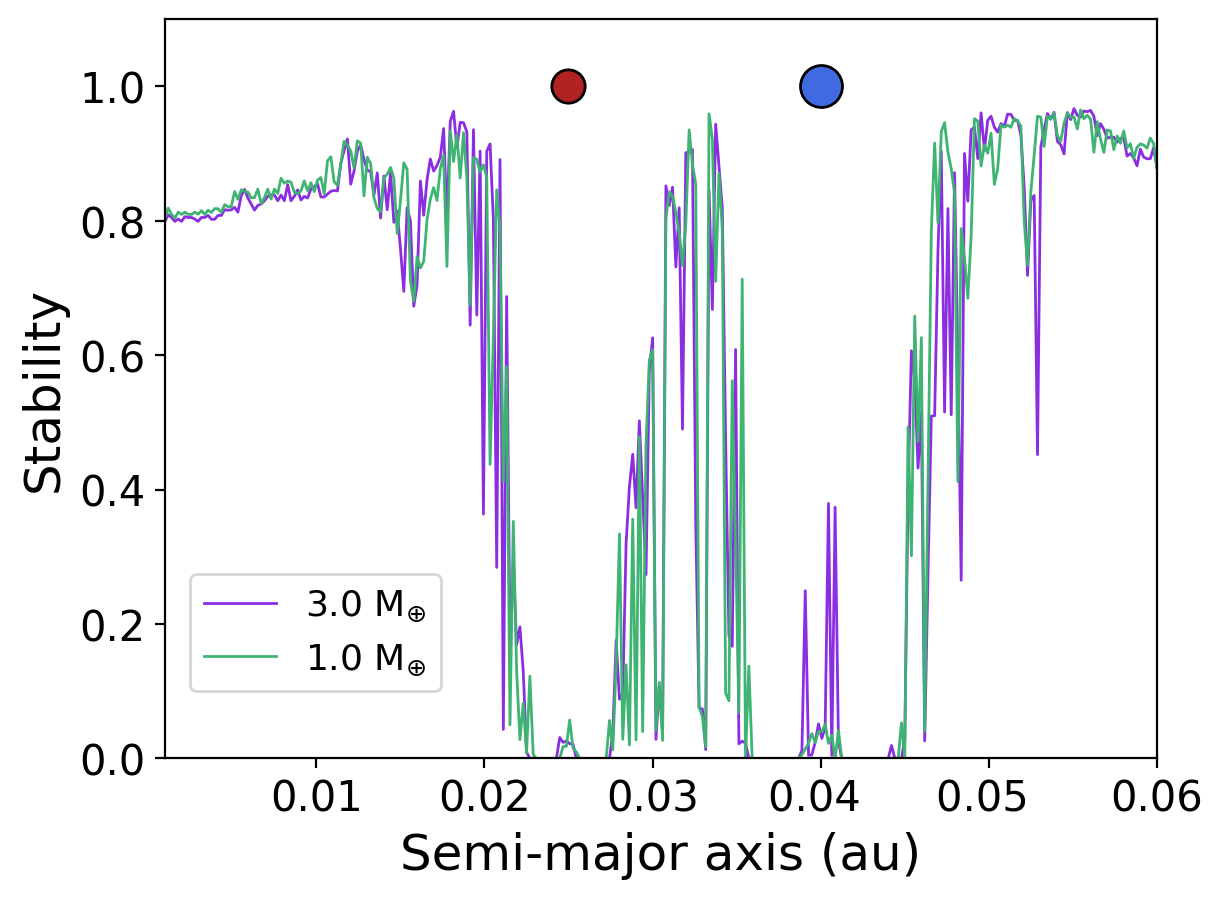}
    \caption{Stability of TOI-2096 as a function of the semi-major axis of a hypothetical third planet injected into the system. The green and the purple lines correspond to an injected mass of 1.0 and 3.0~M$_{\oplus}$, respectively, obtained from the dashed lines in Fig.~\ref{fig:3planet}. The red and the blue dots correspond to planets TOI-2096\,b and c, whose sizes scale with the radii reported in Table~\ref{table:planet_params}.}
    \label{fig:3planet_extension}
\end{figure}

In the first trial, we assumed TOI-2096\,b and c reside in circular orbits. This stability map is displayed in Fig.~\ref{fig:3planet}. We found that the system's stability is not dependent on the mass of the hypothetical third planet but only on its location. The unstable locations are presented in dark-blue hues, while the fully stable regions are depicted in yellow. The horizontal dashed lines represent the particular cases for masses of 1.0 (green) and 3.0~M$_\oplus$ (purple), later presented in Fig.~\ref{fig:3planet_extension}. We found that all planetary masses yielded a similar pattern: a stable region from 0 to 0.02~au controlled by planet b, a mid-region whose stability is influenced by the two planets from 0.027 to 0.035~au, and an outer region beyond 0.047~au driven by planet c. These three stable regions are intercalated by two pits of instability, where the location of the extra planet is forbidden. 

In the second step, we studied how the stability map varies when considering different masses and eccentricities for planets b and c. We found that orbital eccentricities erode the mid-region: the larger the eccentricities, the more unstable the region is. On the other hand, the planetary masses directly influence the instability pits' width: the heavier the planets, the wider the pits are. 

\section{Discussion}\label{sec:disc}

\subsection{Prospects for radial velocity follow-up}\label{sec:radialvelo}

Here, we estimate the required effort to measure the planetary masses using the RV technique. Based on recent studies of small transiting planets orbiting M dwarfs \citep{LuquePalle2022}, TOI-2096\,b and c are expected to be, given their radii of $\sim$1.2~R$_\oplus$ and $\sim$1.9~R$_\oplus$, a rocky planet and a water world, respectively. TOI-2096\,c could also be a rocky planet with this radius, but it would then be the largest known rocky planet orbiting an M dwarf to date. Assuming these bulk compositions, TOI-2096\,b has a predicted mass of $\sim$2~M$_\oplus$, which translates to an RV semi-amplitude of about $2.3\,\mathrm{m\,s^{-1}}$ assuming a circular orbit. For TOI-2096\,c, assuming a water--rich (rocky) composition, the predicted mass is $\sim$4~M$_{\oplus}$ ($\sim$10~M$_{\oplus}$), which gives an RV semi-amplitude of $3.7\,\mathrm{m\,s^{-1}}$ ($9.0\,\mathrm{m\,s^{-1}}$). 
While the predicted semi-amplitudes of the planets make their RV detection feasible with current state-of-the-art RV instruments, the low brightness and high northern declination of the host limit the possibilities of follow-up. With a magnitude in $V$- and $J$-bands of 15.8 and 11.9~mag, respectively, it is out of reach for instruments mounted in 3--4 meter class telescopes in the northern hemisphere such as HARPS-N \citep{HARPN}, CARMENES \citep{CARMENES}, NEID \citep{NEID} or EXPRES \citep{EXPRES}. Near-infrared instruments such as IRD \citep{IRD} or HPF \citep{HPF} mounted in 8--10 meter class telescopes could obtain high-enough S/N to derive RV measurements, but have not demonstrated the internal RV precision below $2\,\mathrm{m\,s^{-1}}$ required to measure the masses of the planets in such a faint system \citep[see, e.g.,][]{2022PASJ...74..904H,2021AJ....162..135K}.

MAROON-X at Gemini North \citep{MAROON-X} is particularly suited to obtain high-precision RVs for M dwarf hosts due to its broad red-optical wavelength coverage. Assuming an exposure time of 30~min and excellent weather conditions, the predicted RV precision achievable on target is about $3\,\mathrm{m\,s^{-1}}$ in the blue arm and $1.2\,\mathrm{m\,s^{-1}}$ in the red arm. Assuming average weather conditions, the photon noise limited RV uncertainties go up to 5 and $3\,\mathrm{m\,s^{-1}}$ in the blue and red arms, respectively. Therefore, it could be possible to measure the masses of the two transiting planets in the TOI-2096 system using MAROON-X. The number of observations required to do so depends on the level of precision desired in the mass determination, the actual internal RV precision achieved on target, and the level of stellar variability present in the spectroscopic measurements. 

It is worth noting that the estimations for the RVs described above refer to the case of circular orbits; however, from our analyses in Sect.~\ref{sec:transits}, we found that eccentric orbits for both planets seem to be favored. In such a case, under the same assumptions regarding the planetary compositions and weather conditions, the induced RVs semi-amplitudes would be slightly larger and hence easier to detect.

\subsection{TOI-2096 and the radius valley}
\label{sec:radiusvalley}

\begin{figure}
    \centering
    \includegraphics[width=0.5\textwidth]{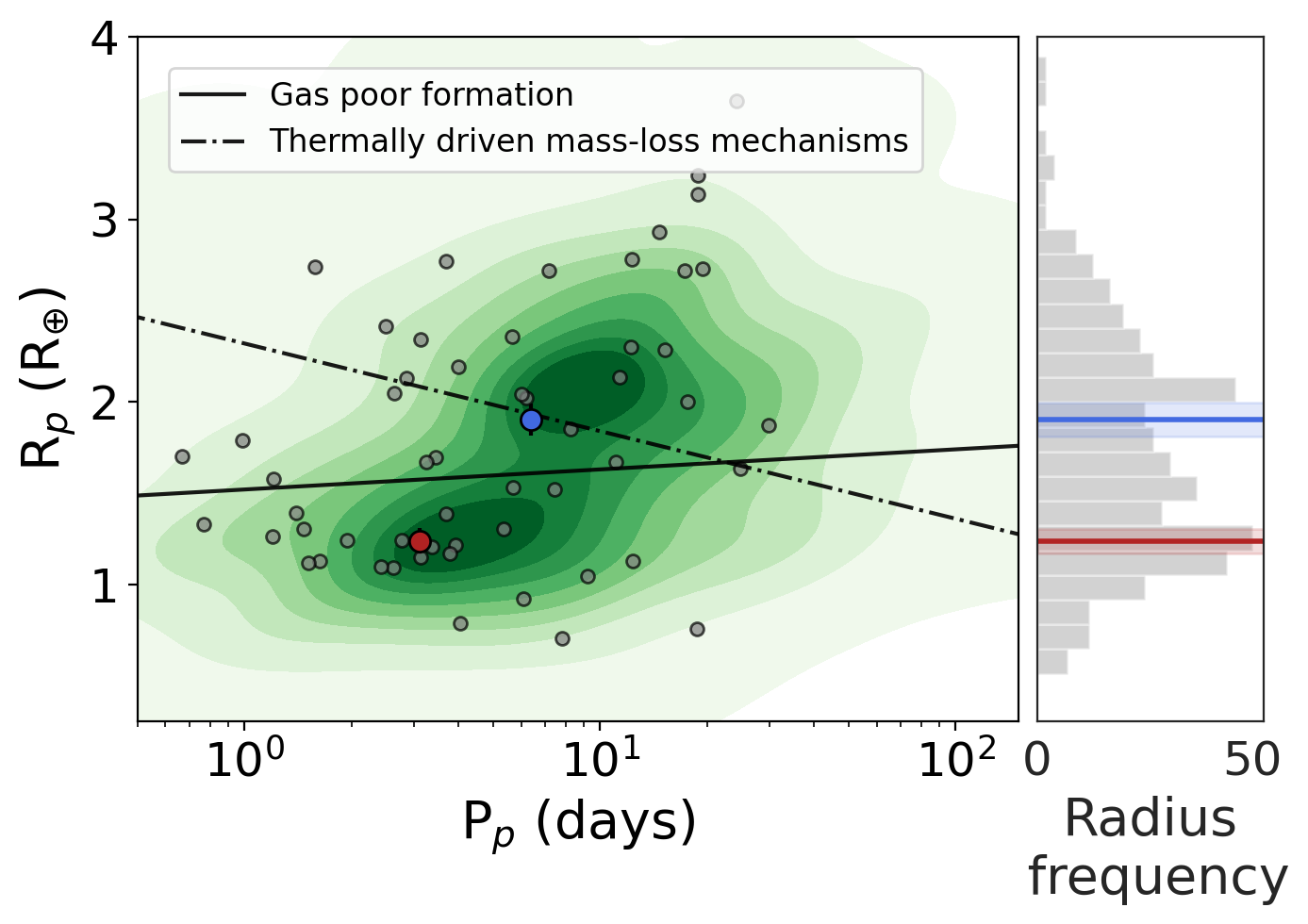}
    \caption{The left panel depicts the planet radius--orbital period plane, containing all the transiting planets with sizes <4~R$_{\oplus}$ orbiting low-mass stars (stellar mass < 0.70 M$_{\odot}$). The green contours represent the density distribution of all of these planets. The gray dots are planets with density estimations better than 30$\%$. The dashed line refers to the prediction using thermally driven mass-loss mechanisms, such as photoevaporation and core-powered mass-loss. On the other hand, the solid line represents the prediction from the gas-poor formation model. 
Planets inside the region limited by these two lines are referred to as `keystone' planets. The red and the blue dots refer to the planets TOI-2096\,b and c, respectively. The right panel displays the radius frequency, showing the radius valley and the location of planets b (red band) and c (blue band) encompassing it.}
    \label{fig:radiusvalley}
\end{figure}

Studies using Kepler planets have shown a paucity of planets with planetary radii from 1.7 to 2.0~R$_\oplus$ around FGK stars 
\citep[see, e.g.,][]{owen2013,fulton2017,fulton2018}, and from 1.4 to 1.7~R$_\oplus$ around low-mass stars from mid-K to mid-M \citep{hirano2018,cloutier2020RV}. This feature is the so-called radius valley, which hints at an orbital-separation-dependent transition between small-rocky and nonrocky planets with extended H/He envelopes \citep{weiss2014,dressing2015}. 
To explain this bimodality, a variety of thermally driven mass-loss mechanisms have been proposed, such as photoevaporation and core-powered mass-loss. The former suggests that gaseous envelopes of small close-in exoplanets might be stripped by XUV radiation from their host stars \citep[see, e.g.,][]{jackson2012,lopez2014,jin2018} on short-time scales of $\sim$100~Myr. On the other hand, the latter proposes that the atmospheric mass loss is a slow process acting over a gigayear timescale, driven by the dissipation of the planetary core's primordial energy from formation \citep{ginzburg2018,gupta2019,gupta2020}. These two mechanisms are both less efficient toward low-mass stars, where the alternative explanation invokes the formation of distinct rocky and nonrocky planet populations in a gas-poor environment \citep{lee2014,lee2016,lopez2018}. The current period--radius diagram of all known exoplanets, with precise radius measurements orbiting M dwarfs, is shown in Fig.~\ref{fig:radiusvalley}. The empirical locations of the radius valley for FGK stars (dashed line) given by  \cite{eylen2018}, and for low-mass stars (solid line) as concluded by \cite{cloutier2020RV} are also displayed. 

The thermally driven mass-loss mechanisms and gas-poor formation models have negative and positive slopes, respectively. These two trends yield a region of interest that offers opposing predictions. According to \cite{cloutier2020RV}, planets residing inside it are valuable targets to conduct tests to elucidate which model dominates the radius valley's emergence across a range of stellar masses. These planets are referred to as "keystones", and their complete characterization, that is, deriving their bulk densities, is highly desired. A very few well-characterized planets, with density precisions better than 30$\%$ lie in this region (see Fig.~\ref{fig:radiusvalley}): TOI-1235\,b \citep{bluhm2020,cloutier2020TOI1235}, TOI-776\,b \citep{luque2021}, TOI-1634\,b \citep{cloutier2021TOI1634,  hirano2021}, TOI-1685\,b \citep{bluhm2021,hirano2021}, GJ~9827\,b \citep{rice2019,dai2019}, TOI-1452\,b \citep{cadieux2022}, TOI-178\,c \citep{leleu2021}, and K2-146\,b \citep{hamann2019,lam2020}. However, these planets do not seem to follow the same evolutionary path. Hence, much more planets with accurate density estimations are needed to robustly state if the radius valley results from atmospheric erosion or the gas-poor formation scenario or a consequence of planet formation and internal composition \citep{LuquePalle2022}.      

In this context, TOI-2096\,b is located in the region where both models predict the location of small, rocky worlds. On the other hand, TOI-2096\,c is located in the area of interest described above (see Fig.~\ref{fig:radiusvalley}). In this study, we cannot determine the planetary masses with the current data set. However, while time intensive, the planetary masses might be obtained using high-precision photometry to accurately measure the TTVs (see Sect.~\ref{sec:dyn_ttv}) or using a RVs follow-up (see Sect.~\ref{sec:radialvelo}). In such a case, the TOI-2096 system, having two planets on the borders of the radius valley, might be an interesting case of study, since a unique evolutionary history of the host star and initial gaseous disk should explain the formation of these two different worlds. 

\subsection{Potential for atmospheric characterization}
\label{sec:atmos}

TOI-2096\,b and\,c are two planets that straddle the radius valley, with the latter likely having a hydrogen envelope. Such worlds are of interest for atmospheric studies, so it is useful to ascertain the relative potential for atmospheric characterization of these planets compared to other similar worlds. To this end, we estimated the typical signal amplitude in transit transmission spectroscopy, $\Delta dF$, using the same approach as for TRAPPIST-1's planets \citep{gillon2016}, which is given by:

\begin{equation}
\begin{split}
\Delta dF &=\frac{2 R_p h_{\rm eff}}{R_\star^2}, {\rm with}\\
h_{\rm eff} &= N_{H} H,
\label{eqn:AtmS}
\end{split}
\end{equation}
where R$_{p}$ is the planetary radius, R$_{\star}$ is the stellar radius, and $h_{eff}$ is the effective atmospheric height (that is, the extent of the atmospheric annulus). We note that $h_{eff}$ is directly proportional 
to the atmospheric scale height given by $H=kT/ \mu g$, where $k$ is the Boltzmann's constant, $\mu$ is the atmospheric mean molecular mass, $T$ is the atmospheric temperature, and $g$ is the surface gravity. The ratio $h_{eff}/H$ for a transparent atmosphere is typically between 6 and 10 \citep{miller2009,wit2013} and depends on the presence of clouds and the spectral resolution and range covered. We assumed to cover seven atmospheric scale heights, $N_{H}$=7, $\mu$=20~amu for planets with R$_{p}$<1.6~R$_{\oplus}$ and $\mu$=2.3~amu for planets with R$_{p}$>1.6~R$_{\oplus}$
\citep[see, e.g.,][]{demory2020,niraula2020,delrez2022}. We assumed the atmospheric temperature to be the equilibrium temperature for a Bond albedo of 0. For the planets with missing masses, we 
estimated $g$ using the model from \cite{chen2017} to imply these masses. We note that any deviation in the real masses from these values will, of course, affect the suitability for atmospheric characterization.

\begin{figure*}
    \centering
    \includegraphics[width=0.70\textwidth]{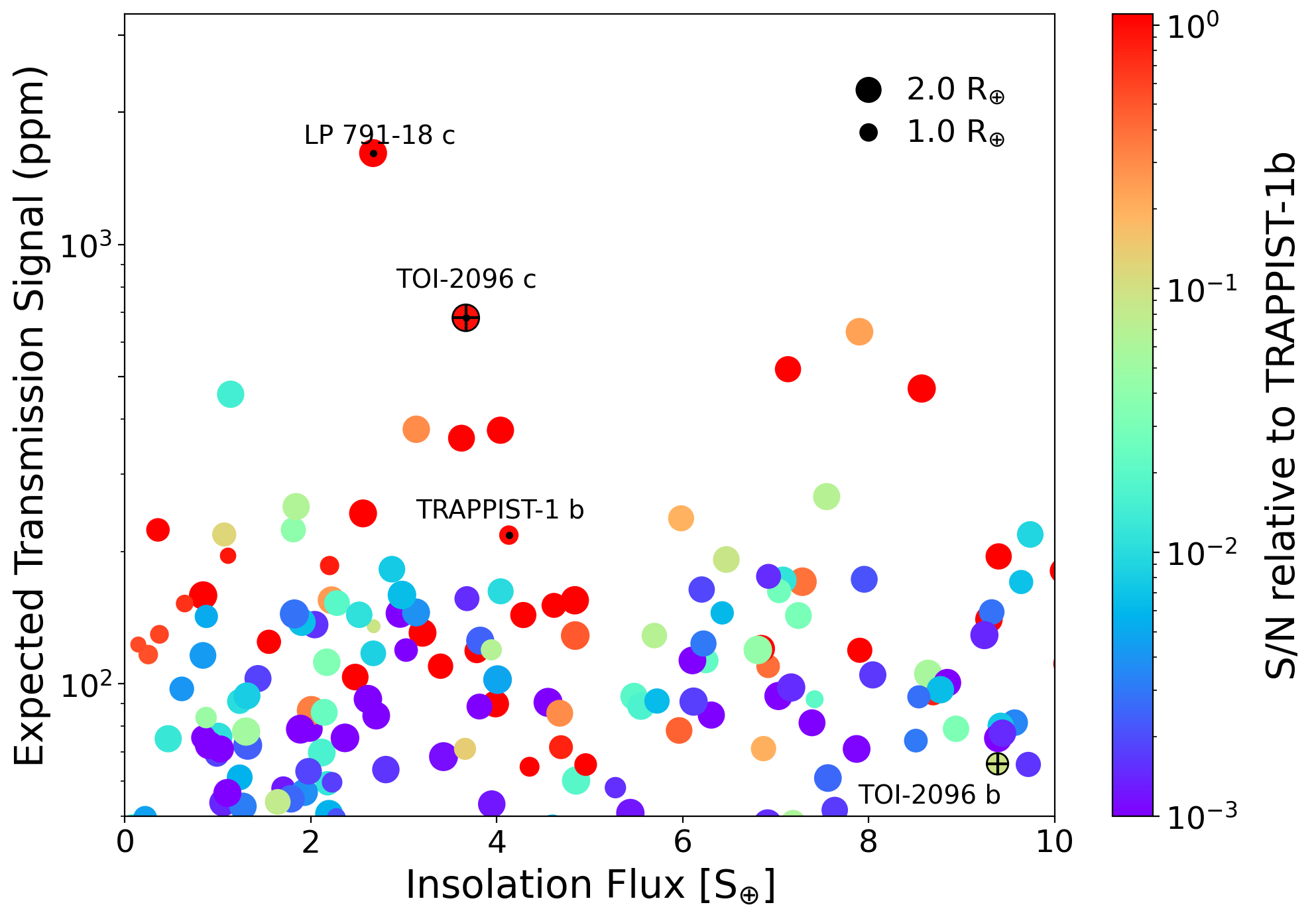}
    \caption{Expected transmission signal as a function of the incident stellar flux for small planets (R$_{p}$<2.5~R$_{\oplus}$). The symbols are color-coded as a function of their S/N relative to TRAPPIST-1\,b, using a JWST/NIRSpec observation. The size of the symbols is proportional to the planetary radius, and circles for 1 and 2 R$_\oplus$ planets are shown for comparison. Planets LP~791-18\,c and TRAPPIST-1\,b are labeled and marked with a central dot for quick recognition. TOI-2096\,b and c are marked with a cross. Data were taken from NASA Exoplanet Archive on October 24, 2022.} \label{fig:atmosphericsignal}
\end{figure*}

Figure~\ref{fig:atmosphericsignal} shows the estimated transmission signal as a function of the planet's incident stellar flux. Each planet is color-coded as a function of its S/N relative to  TRAPPIST-1\,b, obtained by scaling the signal amplitude with the host's brightness in $J$-band and using TRAPPIST-1\, b's S/N as a reference. We restricted the parameter space to planetary sizes <2.5~R$_{\oplus}$ with incident stellar fluxes lower than 10~S$_{\oplus}$, and transmission signals greater than the JWST's threshold of $\sim$50~ppm. We found that in the current picture, TOI-2096\,c stands out as one of the most favorable candidates for atmospheric characterization, with a transmission signal of $\sim$681~ppm. Only LP~791-18\,c \citep{crossfiled2019} has a larger value in this restricted parameter space. On the other hand, for TOI-2096\,b, we found a transmission signal of $\sim$66~ppm, part of the slowly increasing population of likely rocky planets whose atmospheres will be accessible with JWST and future instrumentation, as we discuss in the following subsection. We found the corresponding S/N relative to TRAPPIST-1\,b for TOI-2096\,b and c to be 0.01 and 0.95, respectively.

\subsubsection{James Webb Space Telescope}
\begin{figure}
    \centering
    \includegraphics[width=0.45\textwidth]{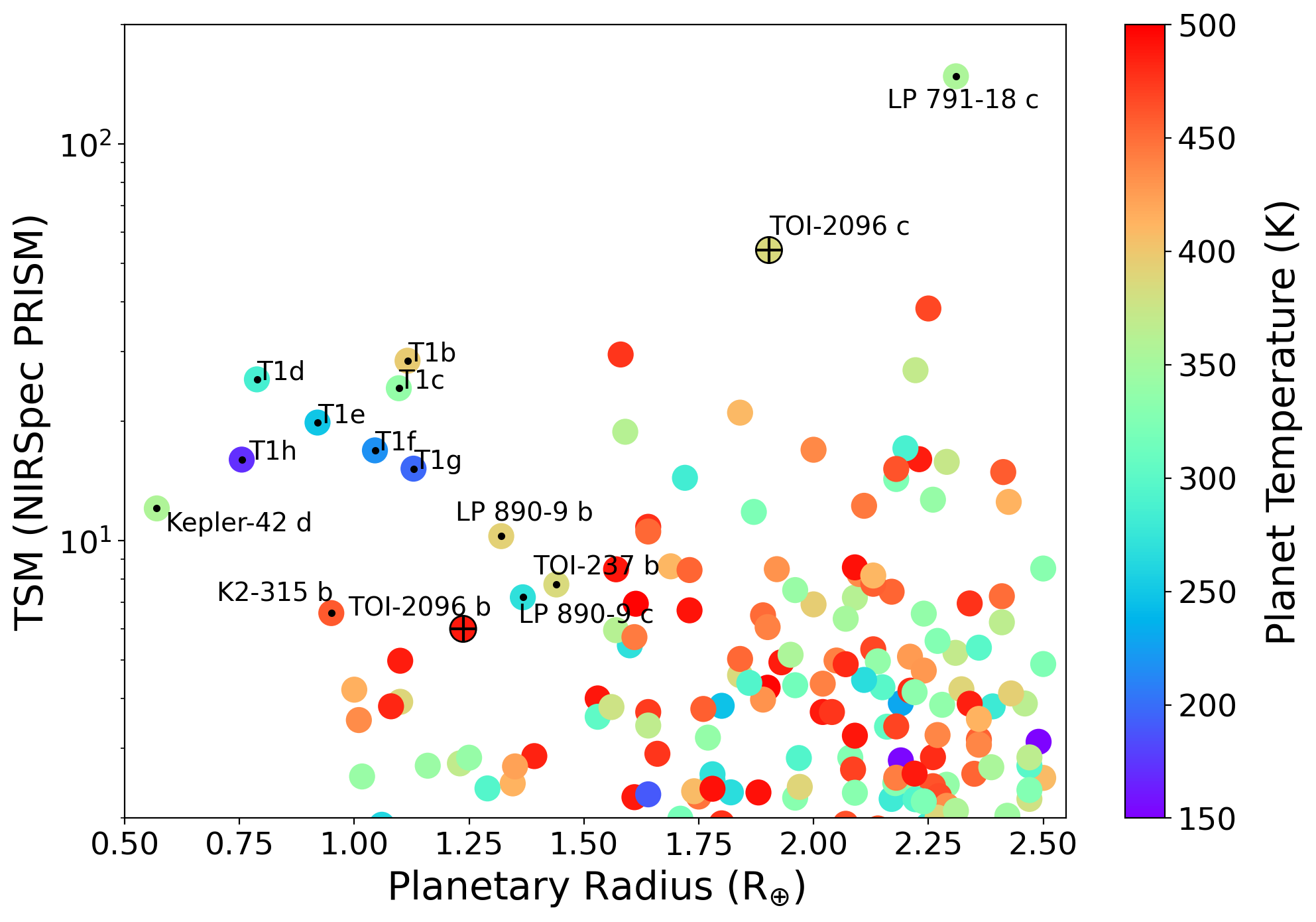}
    \caption{Transmission Spectroscopy Metric for currently known planets, which can be observed with the JWST NIRSpec/Prism. The sample is limited to those with equilibrium temperatures below 500~K and radii smaller than 2.5~R$_{\oplus}$. The color code refers to the planet's temperature. Planets with larger TSM than TOI-2096\,b and c, are labeled and highlighted with a central dot for quick recognition. TOI-2096\,b and c are labeled and highlighted with a cross. Data were taken from NASA Exoplanet Archive on October 24, 2022.     
    } \label{fig:nirspec_prism_tsm}
\end{figure}

In the context of JWST, TOI-2096\,b and c make ideal targets for NIRSpec/Prism (0.6--5.3 $\mu$m) \citep{jakobsen2022}, as this instrument cannot observe stars brighter than J$\sim$10.5 without saturating \citep{birkmann_nirspec} and TOI-2096 has a brightness of J=11.9 (see Table~\ref{table:stellar}). We used the transmission spectroscopy metric \citep[TSM,][]{kempton_tsm} to assess the relative suitability of small (R < 2.5~R$_{\oplus}$), cool (T < 500~K) planets for study with this instrument. As previously, for planets that do not currently have a measured mass, including TOI-2096\,b and c, we used the relation from \citet{chen2017} to estimate it. We found TSM values for TOI-2096\,b and c of 6.0 and 54.0, respectively. Fig.~\ref{fig:nirspec_prism_tsm} displays the TSM values for all the planets in such a parameter space, with TOI-2096\,b and c highlighted. In their class, we find that these worlds are among the best targets for atmospheric characterization with the JWST NIRSpec/Prism. Indeed, for rocky planets with sizes <1.5~R$_{\oplus}$, we found that only a few planets are better suited: all the TRAPPIST-1 planets \citep{gillon2017,agol2021}, Kepler-42~d \citep{kepler42-2012}, LP 890-9\,b and c \citep{delrez2022}, TOI-237\,b \citep{waalkes2021}, and K2-315\,b \citep{niraula2020}. On the other hand, for planetary sizes >1.5~R$_{\oplus}$, only LP 791-18\,c has a larger TSM than TOI-2096\,c. Of course, planets around brighter stars may also be studied by JWST but would require at least two visits to cover the same spectral range as the NIRSpec/Prism (e.g., NIRISS GR700XD \& NIRSepc G395M/H).

\subsubsection{Ariel}
While JWST will conduct many exoplanet observations, it is not an exoplanet-dedicated facility. However, the ESA Ariel mission will observe $\sim$1,000 exoplanets during its 4-year prime mission, conducting a chemical survey of exoplanetary atmospheres \citep{tinetti_ariel, tinetti_ariel2}. Launching in 2029, many of the planets observed by the mission are expected to be found by TESS \citep{edwards_tinetti} and a diverse selection of targets will be observed to search for trends in atmospheric chemistry. 

\begin{figure}
    \centering
    \includegraphics[width=0.45\textwidth]{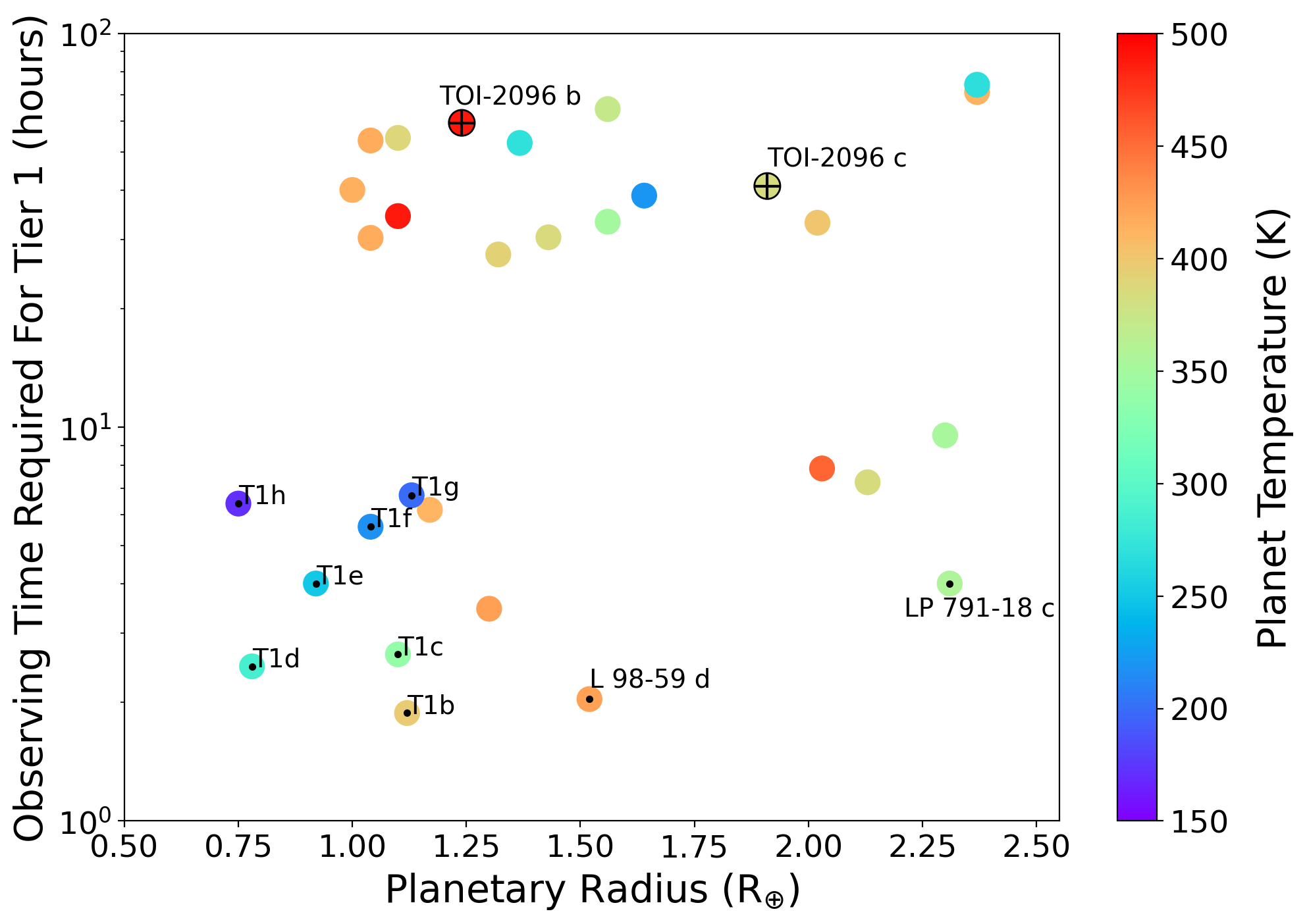}
    \caption{The observing time needed to reach the S/N requirements for Ariel's Tier 1 reconnaissance survey.} \label{fig:ariel_time}
\end{figure}

We evaluated the time required to reach the S/N requirements for Ariel's Tier 1 reconnaissance survey using ArielRad, a radiometric tool developed by the mission's consortium \citep{mugnai_ar}. We find that TOI-2096\,b and\, c would require $\sim$60 and $\sim$40 hours of observing time, respectively. In Fig~\ref{fig:ariel_time}, we compare the time required for these planets to other currently known worlds with similar characteristics. In the context of Ariel, we find that TOI-2096\,b and\, c are no longer the best-in-class as, unlike JWST, the mission can cover a wide spectral coverage (0.5-7.8 $\mu$m) for bright stars. Nevertheless, the parameter space in which TOI-2096\,b and\, c lie is sparsely populated and so they could still be observed by the mission, especially if, during the target selection, emphasis is placed upon studying smaller planets or multiple planets within a single system \citep{edwards_ariel_tl,edwards_tinetti}.

\section{Conclusions} 
\label{sec:concl}

This work presents the discovery and initial characterization of a two-planet system orbiting the nearby M4 dwarf TOI-2096. We validated the planetary nature of TOI-2096\,b and TOI-2096\,c by combining photometry from six TESS sectors and several ground-based facilities, high-resolution imaging, and spectral analysis. We found that the sizes of TOI-2096\,b and c correspond to a super-Earth ($\sim$1.2~R$_{\oplus}$) and a mini-Neptune ($\sim$1.9~R$_{\oplus}$), with orbital periods of 3.12 and 6.39~d, respectively, likely residing in slightly eccentric orbits. We explored the dynamical architecture of the system, where we showed that the derived parameters from our analyses provide long-term stability. Moreover, the system may accommodate an extra planet in a number of close-in orbits without losing its stability.  

The planets' sizes are found to straddle the radius valley. While the size of planet TOI-2096\,b is compatible with a rocky composition, the planet TOI-2096\,c is located in a particular parameter space where different formation models yield different predictions for its composition.  
Their orbital periods are close to the 2:1 MMR, a situation that may allow us to derive their masses through TTVs measurements. While the photometric precision in the current data set prevented us from performing preliminary mass estimations, we found that it is likely achievable by performing a dedicated photometry follow-up with mid-transit-time precisions of $\lesssim$2~min. In addition, we found that, while challenging, the masses of planets may be derived using the RV technique; in particular, the MAROON-X instrument is well suited for such a purpose.   

We also found that in their class, TOI-2096 planets are ideal for atmospheric studies. In particular, using NIRSpec/Prism on board JWST, only a few candidates are better suited, such as the LP\,791-18\,c, and TRAPPIST-1 planets. In the context of the Ariel mission, while the planets are not among the best to be studied using the ArielRad, they lie in a region of the parameter space poorly populated, keeping the door open to be included in dedicated surveys for small planets in multiplanetary systems. These characteristics make the TOI-2096 system appealing for further analyses and studies in various disciplines, such as planetary formation and evolution, multiplanet dynamics, interior modeling, planet-planet and star-planet interactions, comparative planetology, and atmospheric characterization.

\begin{acknowledgements}
We thank the anonymous referee for the helpful comments.
Author F.J.P. acknowledges the support by the Spanish Centro de Excelencia Severo Ochoa (SEV-2017-0709, CEX2021-001131-S).
We acknowledge the use of public TESS data from pipelines at the TESS Science Office and at the TESS Science Processing Operations Center. Resources supporting this work were provided by the NASA High-End Computing (HEC) Program through the NASA Advanced Supercomputing (NAS) Division at Ames Research Center for the production of the SPOC data products. Funding for the TESS mission is provided by NASA’s Science Mission directorate.
This publication makes use of data from the TRAPPIST project. TRAPPIST is funded by the Belgian National Fund for Scientific Research (F.R.S.-FNRS) under grant PDR T.0120.21. TRAPPIST-North was funded by the University of Liège, and is operated in collaboration with the Cadi Ayyad University
of Marrakesh. V.V.G. is a F.R.S.-FNRS Research Associate. M.G. is F.R.S-FNRS Research Director. E.J. is F.R.S-FNRS Senior Research Associate. L.D. is an F.R.S.-
FNRS Postdoctoral Researcher.
The Liverpool Telescope is operated on the island of La Palma by Liverpool John Moores University in the Spanish Observatorio del Roque de los Muchachos of the Instituto de Astrofisica de Canarias with financial support from the UK Science and Technology Facilities Council.
J.d.W. and MIT gratefully acknowledge financial support from the Heising-Simons Foundation, Dr. and Mrs. Colin Masson and Dr. Peter A. Gilman for Artemis, the first telescope of the SPECULOOS network situated in Tenerife, Spain. We thank the staff of the Observatorio del Teide, and especially the "T\'ecnicos de Operaciones Telescopesc\'opicas", for their continuous support to Artemis operations.
Some of the Observations in the paper made use of the High-Resolution Imaging instrument ‘Alopeke. ‘Alopeke was funded by the NASA Exoplanet Exploration Program and built at the NASA Ames Research Center by Steve B. Howell, Nic Scott, Elliott P. Horch, and Emmett Quigley. Data were reduced using a software pipeline originally written by Elliott Horch and Mark Everett. ‘Alopeke was mounted on the Gemini North telescope of the international Gemini Observatory, a program of NSF’s OIR Lab, which is managed by the Association of Universities for Research in Astronomy (AURA) under a cooperative agreement with the National Science Foundation. On behalf of the Gemini partnership: the National Science Foundation (United States), National Research Council (Canada), Agencia Nacional de Investigación y Desarrollo (Chile), Ministerio de Ciencia, Tecnología e Innovación (Argentina), Ministério da Ciência, Tecnologia, Inovações e Comunicações (Brazil), and Korea Astronomy and Space Science Institute (Republic of Korea).
This work has made use of data from the European Space Agency (ESA) mission {\it Gaia} (\url{https://www.cosmos.esa.int/gaia}), processed by the {\it Gaia} Data Processing and Analysis Consortium (DPAC, \url{https://www.cosmos.esa.int/web/gaia/dpac/consortium}). 
Funding for the DPAC has been provided by national institutions, in particular, the institutions participating in the {\it Gaia} Multilateral Agreement.
This research is in part funded by the European Union's Horizon 2020 research and innovation program (grants agreements n$^{\circ}$ 803193/BEBOP), and from the Science and Technology Facilities Council (STFC; grant n$^\circ$ ST/S00193X/1).
B.V.R. thanks the Heising-Simons Foundation for its support.
MNG acknowledges support from the European Space Agency (ESA) as an ESA Research Fellow.
B.E. is a Laureate of the Paris Region fellowship program, which is supported by the Ile-de-France Region and has received funding under the Horizon 2020 innovation framework program and the Marie Sklodowska-Curie grant agreement no. 945298.
R.L. acknowledges funding from the University of La Laguna through the Margarita Salas Fellowship from the Spanish Ministry of Universities ref. UNI/551/2021-May 26, and under the EU Next Generation funds.
This work is based upon observations carried out at the Observatorio Astron\'omico Nacional at the Sierra de San Pedro M\'artir (OAN-SPM), Baja California, M\'exico. We warmly thank the entire technical staff of the Observatorio Astron\'omico Nacional at San Pedro M\'artir for their unfailing support to SAINT-EX operations. YGMC acknowledges support from UNAM-PAPIIT-IG101321.
L.M. acknowledges support from the ``Fondi di Ricerca Scientifica d'Ateneo 2021'' of the University of Rome ``Tor Vergata''. 
J.C.S. acknowledges support by the University of Granada and by Spanish public funds for research under project "Contribution of the UGR to the PLATO2.0 space mission. Phases C / D-1", funded by MCNI/AEI/PID2019-107061GB-C64.
J.D. acknowledges support from the “Alien Earths” program (supported by the National Aeronautics and Space Administration under Agreement No. 80NSSC21K0593) for NASA’s Nexus for Exoplanet System Science (NExSS) research coordination network sponsored by NASA’s Science Mission Directorate.
This material is partly based upon work supported by the National Aeronautics and Space Administration under Agreement No. 80NSSC21K0593 for the program “Alien Earths”. The results reported herein benefited from collaborations and/or information exchange within NASA’s Nexus for Exoplanet System Science (NExSS) research coordination network sponsored by NASA’s Science Mission Directorate. 
P.J.A. acknowledges financial support from the Agencia Estatal de Investigación of the Ministerio de Ciencia e Innovación and the ERDF “A way of making Europe” through projects PID2019-109522GB-C52/AEI/10.13039/501100011033 and the Centre of Excellence “Severo Ochoa” awards to the Instituto de Astrof\'isica de Andaluc\'ia (SEV-2017-0709).

\end{acknowledgements}

\bibliographystyle{aa}
\bibliography{TOI2096.bib}

\clearpage

\onecolumn
\begin{appendix}

\section{Transit fit posterior distributions}

\begin{figure*}[hbt!]
    \includegraphics[width=1.05\textwidth]{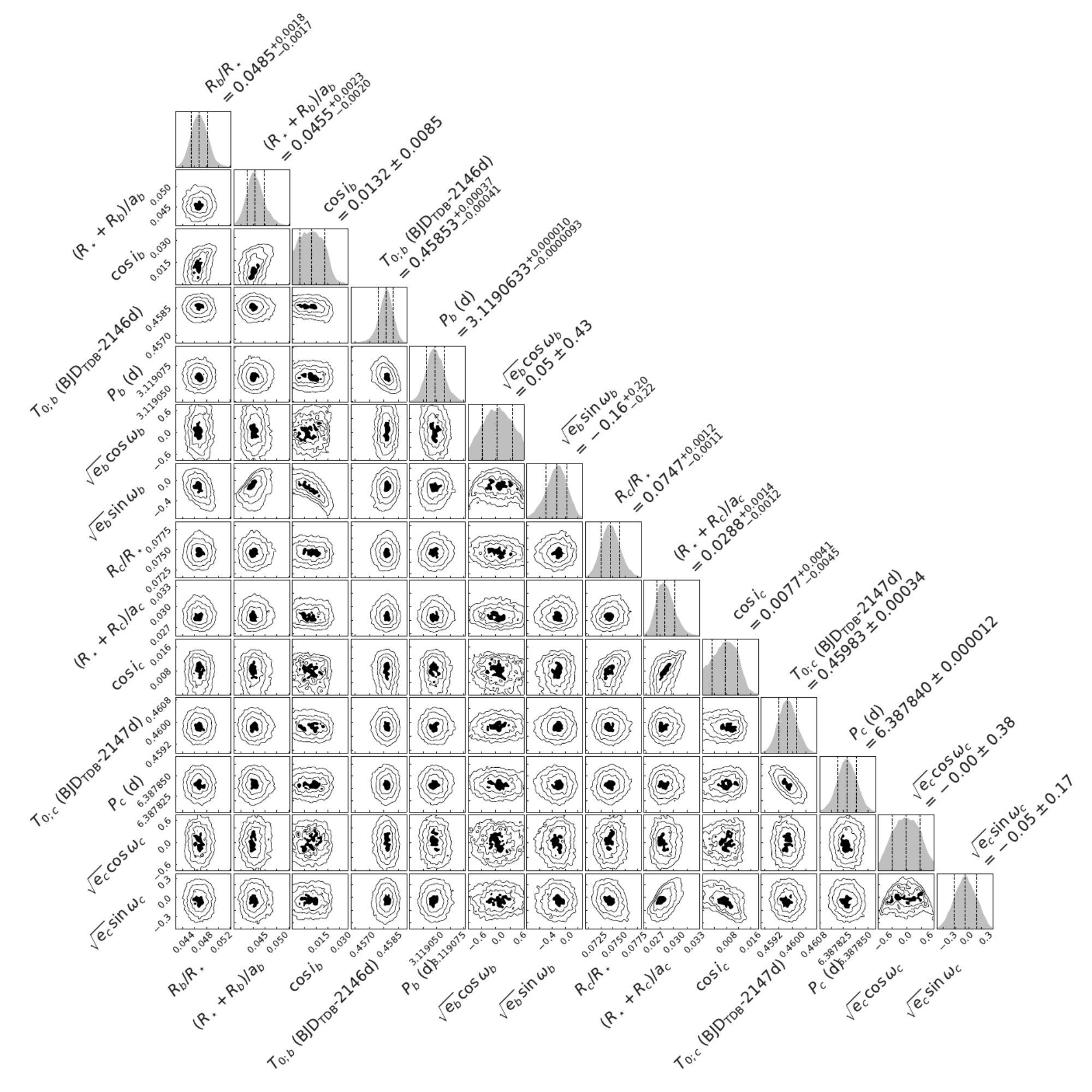}
    \caption{Posterior probability distributions for all the physical parameters fitted using \allesfitter nested sampling as described in Sect.~\ref{sec:transits}. The vertical dashed lines represent the median and the 68$\%$ credible interval. The figure highlights the correlation (or absence thereof) between all the parameters.} 
    \label{fig:posteriors}
\end{figure*}

\clearpage

\section{Quadratic limb darkening coefficients}

\begingroup
\begin{table}[hbt!]
\begin{center}
\renewcommand{\arraystretch}{1.0}
\begin{tabular*}{\linewidth}{@{\extracolsep{\fill}}l c c c c}
\toprule
band & $u_{1}$ & $u_{2}$ & $q_{1}$ & $q_{2}$ \\
\midrule
$z'$ & 0.16$\pm$0.07 & 0.44$\pm$0.04 & 0.36$\pm$0.04 & 0.13$\pm$0.06 \\
$i'$ & 0.31$\pm$0.07  & 0.35$\pm$0.07 & 0.43$\pm$0.06 & 0.23$\pm$0.07  \\
$GG459$ & 0.40$\pm$0.07  & 0.39$\pm$0.04 & 0.63$\pm$0.06  & 0.25$\pm$0.05  \\
$I_{c}$ & 0.27$\pm$0.08 & 0.39$\pm$0.06  & 0.42$\pm$0.06 & 0.20$\pm$0.08  \\
$I+z'$ & 0.21$\pm$0.07 & 0.41$\pm$0.05  & 0.39$\pm$0.04 & 0.17$\pm$0.07 \\
$SDDS$ & 0.16$\pm$0.07 & 0.44$\pm$0.04 & 0.36$\pm$0.04 & 0.13$\pm$0.06 \\
$TESS$ & 0.20$\pm$0.06 & 0.42$\pm$0.04  & 0.38$\pm$0.04 & 0.16$\pm$0.06 \\
\bottomrule
\end{tabular*}
\end{center}
\caption{Quadratic limb darkening coefficients $u_{1}$ and $u_{2}$ from \cite{claret2011}, and their corresponding values $q_{1}$ and $q_{2}$ using the 
relationships from \cite{kipping2013} for each band reported in Table~\ref{tab:GBobservations}.} 
\label{table:LD}
\end{table}
\endgroup

\section{Models comparison}

\begingroup
\begin{table}[hbt!]
\begin{center}
\renewcommand{\arraystretch}{1.15}
\begin{tabular*}{\linewidth}{@{\extracolsep{\fill}}l c c}
\toprule

\multirow{2}{*}{Model} & Free & Bayes factor \\
                       & parameters & $\Delta$ln$Z$ \\
\midrule
Circular orbits for planets b and c & 58 & --  \\
Eccentric orbit for planet b, circular for c & 60  & 4.3  \\
Circular orbit for planet b, eccentric for c & 60  & 3.6  \\
Eccentric orbits for planets b and c & 62 & 4.0  \\
\bottomrule
\end{tabular*}
\end{center}
\caption{Model comparison carried out in Sect.~\ref{sec:transits}. A Bayes factor $>$2.3 would mean strong Bayesian evidence for a model \citep{kass1995} } 
\label{table:model_comparison}
\end{table}
\endgroup

\section{Transit timing variations}
\begingroup
\begin{table}[hbt!]
\begin{center}
\begin{tabular}{c c c c}
\toprule
Predicted timing (BJD$_{TDB}$-2457000) & Observed difference (days) & Date & Telescope \\
\midrule
\multicolumn{4}{c}{\textit{TOI-2096\,b}} \\
\midrule
\smallskip
2099.67258	&  $-0.00046^{+0.00734}_{-0.00319}$  &  07 Sep 2020 & OMM-1.6m\\
\smallskip
2124.62508	&  $-0.00360^{+0.00786}_{-0.00420}$  &    01 Oct 2020 &  TRAPPIST-North-0.6m \\
\smallskip
2177.64916	&  $0.00086^{+0.00151}_{-0.00210}$  &    23 Nov 2020 &  TRAPPIST-North-0.6m \\
\smallskip
2187.00635	&  $-0.00067^{+0.00177}_{-0.00609}$  &    02 Dec 2020 & SAINT-EX-1.0m \\
\smallskip
2205.72073	&  $0.00110^{+0.00079}_{-0.00071}$  &  21 Dec 2020 & Liverpool-2.0m \\
\smallskip
2255.62574	&  $-0.00180^{+0.00108}_{-0.00117}$  &  09 Feb 2021 & Artemis-1.0m \\
\smallskip
2258.74480	&  $-0.00074^{+0.00329}_{-0.00115}$  &  12 Feb 2021 & Artemis-1.0m  \\  
\smallskip
2277.45918	&  $-0.00522^{+0.00138}_{-0.00148}$  &  03 Mar 2021 & TRAPPIST-North-0.6m \\  
\midrule
\multicolumn{4}{c}{\textit{TOI-2096\,c}} \\
\midrule
\smallskip
2134.68415	&  $0.00003^{+0.00090}_{-0.00084}$  &  11 Oct 2020 & TRAPPIST-North-0.6m \\  
\smallskip
2204.95039	&  $0.00040^{+0.00193}_{-0.00123}$  &  20 Dec 2020 & LCO-1.0m \\ 
\smallskip
2230.50175	&  $-0.00089^{+0.00092}_{-0.00083}$  &  15 Jan 2021 & TRAPPIST-North-0.6m \\  
\smallskip
2262.44094	&  $-0.00025^{+0.00070}_{-0.00068}$  &  16 Feb 2021 & Artemis-1.0m \\  
\smallskip
2294.38014	&  $0.00019^{+0.00037}_{-0.00036}$  &  20 Mar 2021 & Artemis-1.0m \\  
\hline 
\end{tabular}
\end{center}
\caption{
Transit timings found in the TTVs analysis conducted in Sect.~\ref{sec:ttv1}
\label{tab:ttvs}}
\end{table}
\endgroup

\end{appendix}

\end{document}